\documentclass[times,authoryear,5p]{elsarticle}
\usepackage{graphicx,amssymb}

\journal{Icarus}
\begin{document}
\begin{frontmatter}
\title{Reassessing the formation of the inner Oort cloud in an embedded star cluster}
\author{R. Brasser\corref{ca}}
\ead{brasser\_astro@yahoo.com}
\address{D\'{e}partement Cassiop\'{e}e, University of Nice - Sophia Antipolis, CNRS, Observatoire de la C\^{o}te d'Azur; Nice,
France}
\cortext[ca]{Corresponding author}
\author{M. J. Duncan}
\ead{duncan@astro.queensu.ca}
\address{Department of Physics, Engineering Physics and Astronomy, Queen's University; Kingston, ON, Canada}
\author{H. F. Levison}
\ead{hal@boulder.swri.edu}
\address{Department of Space Studies, South West Research Institute; Boulder, CO, USA}
\author{M. E. Schwamb}
\ead{megan.schwamb@yale.edu}
\address{Department of Physics and Yale Centre for Astronomy and Astrophysics, Yale University; New Haven, CT, USA}
\author{M. E. Brown}
\ead{mbrown@caltech.edu}
\address{Division of Geological and Planetary Sciences, California Institute of Technology; Pasadena, CA, USA}
 
\begin{abstract}
We re-examine the formation of the inner Oort comet cloud while the Sun was in its birth cluster with the aid of numerical
simulations. This work is a continuation of an earlier study (Brasser et al., 2006) with several substantial modifications.
First, the system consisting of stars, planets and comets is treated self-consistently in our N-body simulations, rather than
approximating the stellar encounters with the outer Solar System as hyperbolic fly-bys. Second, we have included the expulsion of the
cluster gas, a feature that was absent previously. Third, we have used several models for the initial conditions and density profile of
the cluster -- either a Hernquist or Plummer potential -- and chose other parameters based on the latest observations of embedded
clusters from the literature. { These other parameters result in the stars being on radial orbits and the cluster collapses.} Similar to
previous studies, in our simulations the inner Oort cloud is formed from comets being scattered by Jupiter and Saturn and having their
pericentres decoupled from the planets by perturbations from the cluster gas and other stars. We find that all inner Oort clouds formed in
these clusters have an inner edge ranging from 100~AU to a few hundred AU, and an outer edge at over 100\,000~AU, with little variation in
these values for all clusters. All inner Oort clouds formed are consistent with the existence of (90377) Sedna, an inner Oort cloud dwarf
planetoid, at the inner edge of the cloud: Sedna tends to be at the innermost 2\% for Plummer models, while it is 5\% for Hernquist models.
We emphasise that the existence of Sedna is a generic outcome. We define a `concentration radius' for the inner Oort cloud and find that its
value increases with increasing number of stars in the cluster, ranging from 600~AU to 1500~AU for Hernquist clusters and from 1500~AU to
4000~AU for Plummer clusters. The increasing trend implies that small star clusters form more compact inner Oort clouds than large
clusters. We are unable to constrain the number of stars that resided in the cluster since most clusters yield inner Oort clouds that could
be compatible with the current structure of the outer Solar System. The typical formation efficiency of the inner Oort cloud is 1.5\%,
significantly lower than previous estimates. We attribute this to the more violent dynamics that the Sun experiences as it rushes through
the centre of the cluster during the latter's initial phase of violent relaxation.
\end{abstract}
\end{frontmatter}
\section{Introduction and background}
There have been several studies of Oort cloud formation in a star cluster environment. In 1999 Soenke Eggers published his Ph.D.
thesis in which he analysed the effects of a star cluster on the formation and evolution of the Oort cloud (OC). He used a Monte Carlo
method and two star clusters, in which the stellar encounters occurred at constant time intervals and their effects on the comets were
computed analytically. The first cluster had an effective { number} density of 625 stars pc$^{-3}$ and the other had an effective {
number} density two orders of magnitude lower. For both clusters Eggers assumed a velocity dispersion of 1~km~s$^{-1}$. His model did not
include a tidal field caused by the cluster potential. Eggers defined a comet to be in the Oort cloud if it attained $q > 33$~AU and
simultaneously $a > 110$~AU. With these definitions, he obtained efficiencies of 1.7\% and 4.8\% for the low and high density clusters
respectively, with comets on orbits with a typical semi-major axis between 3\,000~AU and 6\,000~AU. The clouds were also found to be mostly
isotropic. \\

A parallel study of the formation of the inner Oort cloud in a denser, clusteresque environment has been performed by Fern\'{a}ndez \&
Brun\'{i}ni (2000). They placed comets on eccentric orbits with semi-major axes $\sim 200$~AU and included an approximate model of the
tidal field of the gas and passing stars from the cluster. The cluster had a maximum stellar { number} density of 100~pc$^{-3}$ and the
maximum { mass} density of the core of the molecular cloud gas was 5000~$M_{\odot}$~pc$^{-3}$. Their simulations formed a dense inner
Oort cloud where the comets had semi-major axes of a few hundred to a few thousand astronomical units. The outer edge of this cloud was
dependent on the density of gas and stars in the cluster. The most interesting part of their study was their ability to successfully deposit
a fair amount of material that was scattered by Jupiter and Saturn, which were the main contributors to forming the inner Oort cloud. In the
current environment, on the other hand, the contribution to the Oort cloud from Jupiter and Saturn is lower than that from Uranus and
Neptune (e.g. Dones et al., 2004). However, Fern\'{a}ndez \& Brun\'{i}ni (2000) pointed out that if the Sun remained in this dense
environment for long, the passing stars could strip a significant fraction of the comets away from the Sun and the trapping efficiency
might end up being low. { This low trapping efficiency was partially a result of their cluster lifetimes being too long.}\\

The interest in the formation of the inner Oort cloud in a cluster environment gained renewed interest with the discovery of (90377)
Sedna, a dwarf planet with semi-major axis 500~AU and perihelion of 76~AU, so that its orbit is detached from that of Neptune (Brown et
al., 2004). Gladman et al. (2002) has shown that the object 2000 CR$_{105}$, having an orbit with $q=44$~AU and $a \sim 200$~AU, could
not be reproduced via chaotic diffusion, which ceases beyond $q=38$~AU. Thus another mechanism had to be responsible { for placing both
2000 CR$_{105}$ and Sedna on their current orbits}. Morbidelli \& Levison (2004), together with Kenyon \& Bromley (2004), { successfully}
demonstrated that the most viable way to reproduce the orbits of Sedna and 2000 CR$_{105}$ is via a slow, close passage of a relatively
heavy star. Morbidelli \& Levison (2004) argued that the encounter had to happen early in order to still have a reasonably populated Oort
cloud afterwards. However, the low velocity of the encounter is difficult to obtain in the current Galactic environment, and they suggested
that this passage occurred while the Sun was in its birth cluster. \\

The above results led to Brasser et al. (2006) - henceforth BDL6 - to investigate the formation of the inner Oort cloud in an embedded
cluster environment, in which the gas from the molecular cloud is still present (Lada \& Lada, 2003). Inspired by Fern\'{a}ndez \&
Brun\'{i}ni (2000) they attempted to constrain the environment that was needed to save comets under the dynamical control of Jupiter.
They employed a Plummer model (Plummer, 1911) to construct a series of clusters with varying central density but with a more or less
fixed number of stars, because most stars form in clusters of a few hundred stars (Lada \& Lada, 2003). BDL6 used a simple
leapfrog integrator for the cluster in the Plummer potential and recorded the positions and velocities of stars, including time, as
they came within a user-specified distance of the Sun. The encounter data were then used in SWIFT RMVS3 (Levison \& Duncan, 1994).
The latter was modified to include the effects of the stars, as in Dones et al. (2004), and the gravitational force from the cluster
gas on the comets and the planets. The Sun was assumed to be on a fixed orbit in the Plummer potential from which the tidal torque of
the gas on the comets was computed. The typical efficiency for the formation of the inner Oort cloud was 10\%. They showed
that Sedna's orbit could be reproduced when the central density of the cluster exceeded 10\,000~$M_{\odot}$~pc$^{-3}$ in gas and stars.
For these clusters Sedna was found to be at the inner edge of the inner Oort cloud. In order to reproduce the orbit of 2000
CR$_{105}$, an even higher (central) density was needed. However, the orbital distribution of the inner Oort cloud in these very high
density clusters is found to be inconsistent with the current observations of the outer solar system (Schwamb et al., 2010), so
that the clusters where Sedna is at the inner edge are preferred.\\

In a similar study, Kaib \& Quinn (2008) studied inner Oort cloud formation in an open cluster with stellar { number} densities ranging
from 10~pc$^{-3}$ to 100~pc$^{-3}$. { Kaib \& Quinn (2008) modelled the effect of the stars using the same} approach as Dones et
al. (2004) and BDL6. { The maximum cluster life time was set to 100~Myr and the density of the stars in the cluster decayed linear with
time, to account for mass loss by mutual scattering of the stars. When the cluster had completely disappeared Kaib \& Quinn (2008)
continued their Oort cloud simulations until the age of the Solar System, something BDL6 did not do. Quantitatively their results were
similar to BDL6 and they were able to reproduce Sedna when the stellar number density exceeded 30~pc$^{-3}$.}\\

In a recent publication attempting to solve some of the outstanding problems associated with the Oort cloud as a whole, Levison et
al. (2010) investigated the capture by the Sun of comets from other stars. They simulated embedded clusters ranging from 30 to 300
stars with a star formation efficiency of 10\% to 30\%. They placed a disc of comets around each star with random orientation and orbits
with $q=30$~AU and semi-major axes ranging from 1\,000 to 5\,000~AU. The whole system of stars and comets was simulated until the
median spacing between the stars became 500\,000~AU. From these numerical simulations Levison et al. (2010) concluded that the capture
efficiency is high enough to obtain the current population of the Oort cloud provided that most of the stars in the cluster contained a
similar number of comets to the Sun. { At least} 90\% of the comets in the Oort cloud could be extrasolar in origin.\\

Unfortunately, apart from the Levison et al. (2010) study, which relied on extrasolar comets rather than indigenous comets to
populate the Oort cloud, all of the above works suffer from the limitation that the stars are treated as hyperbolic encounters and the
Sun is on a fixed orbit in the cluster. Both of these assumptions are wrong: the stars' motion with respect to the Sun reverse
direction when their distance to the Sun is of the same order as the size of the current Oort cloud. If the stellar { number} density in
a cloud is $n_*$~pc$^{-3}$, then their average nearest-neighbour distance is $r=0.62n_*^{-1/3}$~pc, and taking a typical density of
$n_*=30$~pc$^{-3}$ yields $r=0.2$~pc or 42\,000~AU. Secondly, mutual scattering among the stars changes their orbits and some end up on
highly elliptical orbits on their way to being ejected. Hence the assumption of a static orbit is no longer valid. A third issue is gas
removal. BDL6 stopped their simulations after 3~Myr by assuming that at this time the Sun left the cluster. Kaib \& Quinn (2008)
decreased the density of their fictitious open cluster linearly and it was gone after 100~Myr. Levison et al. (2010) made the gas go
away exponentially with an e-folding time of 10\,000 yr. Fourth, the BDL6 study relied on very high central gas densities in
order to torque comets under the dynamical control of Jupiter into the inner Oort cloud before they were ejected. While BDL6 argued
that the densities that were chosen are in agreement with the peak densities observed in some embedded clusters (Gutermuth et al.,
2005), their initial conditions can be improved by using more recent observational data and better models for the embedded clusters.
Any reasonable model of the formation of the inner Oort cloud in a star cluster environment has to take the above issues into
account. The aim of this paper, therefore, is to re-investigate the formation of the inner Oort cloud in a cluster environment using
(i) a better model for the embedded star cluster which best matches the current observations, in particular the initial conditions for
the stars, gas and the gas dispersal, and (ii) a computer code that can handle stars, planets and comets at the same time so that there
is no need to rely on the assumption of hyperbolic stellar encounters. \\

Levison et al. (2010) used a computer code, based on SyMBA (Duncan et al., 1998), that was able to integrate both the comets and stars
symplectically and self-consistently without relying on assumptions of hyperbolic flybys. In this study we shall use their code. This
paper is divided as follows. In Section 2 we summarise some of the basic properties of embedded star clusters that we need for our
simulations as inferred from observations. Section 3 deals with the initial conditions and methods of our numerical simulations. In Section
4 we present the properties of the inner Oort cloud resulting from the numerical simulations. In Section 5 we compare these results with
recent observations of the outer Solar System. In Section 6 we present our discussion, followed by the conclusions in Section 7.

\section{Cluster properties and models}
In this section we discuss the models and parameters that we employed for the simulation of the embedded star clusters.

\subsection{Cluster size}
We use the data of embedded star clusters within 2 kpc of the Sun from Lada \& Lada (2003) and Gutermuth et al. (2009). Adams et
al. (2006) use the Lada \& Lada (2003) data and find that most embedded clusters have between 100 to 1\,000 members. The cumulative
distribution of the number of stars, $N$, has a best fit $f=0.637\log N -1.045$ i.e the distribution increases logarithmically with
$N$ (Adams, 2010). \\

\begin{table}
\begin{tabular}{ccccccc}
Name & $R_H$ [pc] & $R_c$ [pc] & $R_{LL}$ [pc] & $N_T$ & $N_C$ & $N_{LL}$ \\ \hline \\
Mon R2 & 2.88 & 2.01 & 1.85 & 235 & 132 & 371\\
IC 348 & 2.1 & 0.55 & 1.0 & 160 & 56 & 300\\
NGC 1333 & 1.19 & 0.51 & 0.49 & 133 & 96 & 143 \\
GGD 12--15 & 2.12 & 0.64 & 1.13 & 119 & 78 & 134 \\
S 106 & 4.16 & 1.26 & 0.3 & 79 & 36 & 160 \\
MWC 297 & 0.92 & 0.39 & 0.5 & 23 & 10 & 37
\end{tabular}
\caption{Clusters common to Gutermuth et al. (2009) and Lada \& Lada (2003) for which the radius is known. The columns are: name, total
radius from Gutermuth et al. (2009), core radius from Gutermuth et al. (2009), radius from Lada \& Lada (2003), total number of stars
from Gutermuth et al. (2009), number of stars in the core from Gutermuth et al. (2009) and number of stars listed in Lada \& Lada
(2003).}
\label{GLL}
\end{table}

\begin{figure}
\resizebox{\hsize}{!}{\includegraphics[angle=-90]{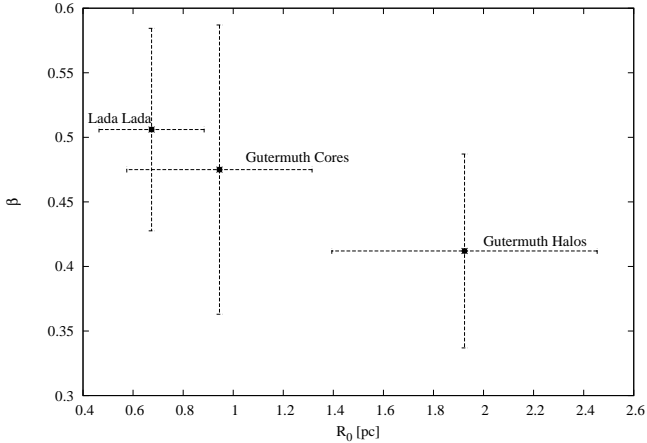}}
\caption{Range in sizes, $R$, vs exponents, $\beta$, for the cluster core sizes from Lada \& Lada (2003) and Gutermuth et al.
(2009).}
\label{fits}
\end{figure}

For most embedded clusters, the observed surface density is a constant (Allen et al., 2007) i.e. $N/R^2$ is constant, where
$R$ is the radius of the cluster. This relation is in agreement with the observed density structure of giant molecular clouds (Blitz
et al., 2007) and theoretical modelling of cloud collapse (Larson, 1985). The observed constant surface density implies that the size
of the cluster scales with the number of stars as $R=R_0N^{1/2}$, where $R_0$ is a scaling parameter. From the catalogue of Lada \&
Lada (2003), Adams et al. (2006) find a best fit $R=R_0(N/100)^{1/2}$~pc where $R_0 \in (0.577,1)$~pc. For their subsequent
simulations Adams et al. (2006) use $R_0=0.577$~pc to maximise dynamical interactions among the stars. However, the radii of embedded
clusters presented in Lada \& Lada (2003) might only be valid for the cores of the clusters that are listed. Recently Gutermuth et al.
(2009) performed a {\it Spitzer} study of a large sample of embedded clusters, some of which are also listed in Lada \& Lada (2003).
Gutermuth et al. (2009) characterise each cluster by a core and an extended 'halo'. Using nearest-neighbour distance counts and
assigning a radius to the cluster as being half the distance between the two farthest stars, a best fit through their data for the
halos yields $R_H=R_0(N/100)^{\beta}$ { where $R_0 =1.92 \pm 0.52$~pc and the exponent $\beta = 0.41 \pm 0.09$.} For the cores
$R_c=R_0(N/100)^{\beta}$ with { $R_0 =0.95 \pm 0.36$~pc and $\beta = 0.47 \pm 0.11$.} The fitting parameters and their error bars are
displayed in Fig.~\ref{fits}. The common clusters of Gutermuth et al. (2009) and Lada \& Lada (2003) for which radii are available are
listed in Table~\ref{GLL}. The columns are: name, total radius from Gutermuth et al. (2009), core radius from Gutermuth et al. (2009),
radius from Lada \& Lada (2003), total number of stars from Gutermuth et al. (2009), number of stars in the core from Gutermuth et al.
(2009) and number of stars listed in Lada \& Lada (2003). Clusters which are common to both catalogues but have no radius listed in Lada \&
Lada (2003) are AFGL 490, IC 5136, Cep C, LH H$\alpha$ 101, Serpens, Cep A, L988-e and R CrA. As can be seen from Table~\ref{GLL}, most of
the sizes listed in Lada \& Lada (2003) are close to the size of the cores listed in Gutermuth et al. (2009), or are intermediate between
the core and the halo. In any case, the cluster sizes in Lada \& Lada (2003) are systematically smaller than in Gutermuth et al. (2009), and
the fit through the core data of Gutermuth et al. (2009) is compatible with the fit through the data of Lada \& Lada (2003), but the halos
are not. \\

The best fits seem to indicate that $R \sim N^{\beta}$ with $\beta \in [1/3,1/2]$. However, the $N^{1/2}$ relation is based on the
assumption that the column or surface density of stars is more or less constant and is an artefact of the way the stars are counted.
Most star identification algorithms rely on the density-weighted nearest-neighbour method of Casertano \& Hut (1985), or the minimum
spanning tree method (e.g. Graham \& Hell, 1985). Both are often employed to identify star clusters (Bastian et al., 2007, 2009; Cartwright
\& Whitworth, 2004; Gutermuth et al., 2009; Schmeja \& Klessen, 2006). However, in all cases the cluster radius is defined as the radius of
a circle with the same area as the projected cluster (Schmeja, 2011). All methods truncate the size of the cluster when the projected
distance between two neighbouring stars is larger than some threshold value, which is equivalent to assuming that the surface density inside
said circle is more or less constant. Thus, the size of the cluster then obviously scales as $N^{1/2}$. However, the $N^{1/2}$ relation
appears in disagreement with another observation, and that is that the average stellar { number} density in these clusters is more or
less constant (Carpenter, 2000; Lada \& Lada, 2003; Proszkow \& Adams, 2009), with a median { number density} of $n_M = 65$~pc$^{-3}$.
From the clusters listed in Lada \& Lada (2003) and Gutermuth et al. (2009), we compute the median value { for the halos of Gutermuth et
al. (2009) to be $n_M = 3.1$~pc$^{-3}$ while for the cores it is $n_M = 46.2$~pc$^{-3}$, comparable to the value listed earlier. However,
the average stellar number density from one cluster to the next can vary by approximately an order of magnitude. This true for both cores
and halos. Thus the values quoted above should be interpreted as indicative only.} \\

It is easy to verify whether the observation of more or less constant stellar { number} density is consistent with the relation between
the cluster's size and the number of stars. The total number of stars in the cluster, apart from a constant, is $N=\langle n \rangle R^3$,
where $\langle n \rangle$ is the average stellar number density. Substituting $R=N^{\beta}$ we have $\langle n \rangle = N^{1-3\beta}$,
which is a constant only if $\beta=1/3$. Thus it seems sensible to adopt a cluster size that scales as $N^{1/3}$. In order to determine
whether or not this scaling makes sense, we turn to observations of open and globular star clusters for guidance. \\

King (1962) proposed that the size of a star cluster, whether open or globular, is given by its tidal radius, $r_t$. At this distance
from the centre, the tidal effects of the Milky Way Galaxy start to dominate over the self gravity of the cluster. King (1962) states
that

\begin{equation}
 r_t = \Bigl(\frac{\mathcal{G}M}{4A(A-B)}\Bigr)^{1/3},
\label{rt}
\end{equation}
where $\mathcal{G}$ is the gravitational constant, $M$ is the total mass of the cluster and $A$ and $B$ are the Oort constants (e.g.
Binney \& Tremaine, 1987). The tidal radius scales as $N^{1/3}$ because $M \propto N$. When adopting a flat Galactic rotation curve
with angular velocity $\Omega_G = 30$~km~s$^{-1}$~kpc$^{-1}$ (MacMillan \& Binney, 2010), we have $A=|B|=15$~km~s$^{-1}$~kpc$^{-1}$ and
so $r_t \sim 4.6\,(N/100)^{1/3}$~pc, which is much larger than the halo sizes of Lada \& Lada (2003) and Gutermuth et al. (2009). This
discrepancy is most likely caused by the fact that for embedded clusters the background density is not that of the Galactic disc, but rather
of the surrounding molecular cloud. Typical densities of molecular clouds are some $\sim 1$~M$_{\odot}$~pc$^{-3}$, so that the tidal radius
in equation (\ref{rt}) above should be divided by $\sim 2$ i.e. $r_t \sim 2.3\,(N/100)^{1/3}$~pc. This result is similar to and compatible
with the halo distances obtained by Gutermuth et al. (2009). The shape of the zero-velocity curves of the cluster reduce { the tidal
radius} even further (Innanen et al., 1983) to $r_t \sim 1.7\,(N/100)^{1/3}$~pc. Given the uncertainties in the observed properties of the
clusters and in the size versus number of stars, we shall anchor the value of $r_t$ for $N=100$ to the best-fit halo value of Gutermuth et
al. (2009) and thus use $r_t=1.92\,(N/100)^{1/3}$~pc for the size of the cluster. \\

\subsection{Core radius and internal structure}
Open and globular star clusters usually show a dense core with more or less constant surface brightness and then an extended halo
where the surface density falls off (King, 1962), typically as $r^{-\gamma}$, where $\gamma \in (2,3)$ (King, 1962; Elson et al.,
1987). Observers generally define the cluster core radius, $r_c$, as the distance from the cluster centre at which the surface
brightness drops by a factor of two from the central value. Theorists, however, often use a definition based on the central density,
$\rho_0$, and central velocity dispersion, $\sigma_0$. The core radius is then given by (King, 1966)

\begin{equation}
r_c = \frac{3\sigma_0}{2\sqrt{\pi G \rho_0}},
\end{equation}
which we shall use here. For most cluster models the core radius corresponds roughly to where the stellar volume density has
decreased by a factor of 3. A second method computes a local density using a star's nearest neighbours (Casertano \& Hut 1985), in
which the core radius becomes a density-weighted quantity obtained from the root-mean-square stellar distances. We refer to Portegies
Zwart et al. (2010) for a more in-depth discussion on how the core radius is defined and measured. King (1962) defines the
'concentration ratio' of the cluster as $c_K = \log (r_t/r_c)$, where the log is with base ten; we shall use the non-logarithmic form
$c=r_t/r_c$ here. For young open clusters (Piskunov et al., 2008) the value of $c$ is typically 3 to 6. The embedded cluster table of
Gutermuth et al. (2009) yields similar values. For the young open, non-relaxed cluster NGC 6611, whose estimated age is 1.3~Myr, $c \sim 10$
(Bonatto et al., 2006). All of these observations suggest that $c$ ranges from approximately 3 to 10, and thus these clusters have fairly
shallow profiles and central potential wells. We shall use $c=3$ and $c=6$ in our simulations.\\

Inside the cluster the volume density scales as $\rho(r) \propto r^{-\gamma}$ where $\gamma \in (0,2)$ (Schmeja \& Klessen,
2006; Schmeja et al., 2008; Andr\'{e} et al., 2007), with values in the range 0 to 1 being the most common. It is well known that the
density of most globular clusters are best fitted with King profiles (King, 1966), which have a well-defined core and halo. Inside the
core the density is more or less constant while outside the core the density falls off quickly (Portegies Zwart et al., 2010). However,
it is unclear if the King profiles are suitable for young/embedded clusters (Portegies Zwart et al., 2010). Since the potential for
the King models cannot be written in closed form as a function of distance from the centre, which we need in our computer code, we
prefer not to use these models. There exist various alternatives in the literature to compute the density and potential. For spherical
Galaxies the models by Dehnen (1993) and Tremaine et al. (1994) are often used. The volume density of these profiles
are

\begin{equation}
 \rho(r) = \frac{(3-\gamma)M}{4\pi} \frac{a_D}{r^\gamma(r+a_D)^{4-\gamma}},
\end{equation}
where $M$ is the total mass in gas and stars, $a_D$ is a parameter radius and $\gamma$ measures the density concentration at the
centre. The density profiles of Jaffe (1983) and Hernquist (1990) are the cases with $\gamma=2$ and $\gamma=1$ respectively. In their
cluster simulations Adams et al. (2006) use the density profile of Hernquist (1990), for which the density is given by

\begin{equation}
 \rho_H(r) = \frac{M}{2\pi} \frac{a_H}{r}\frac{1}{(r+a_H)^3},
\end{equation}
where $a_H$ is the Hernquist radius. The corresponding potential is

\begin{equation}
 \Phi_H(r) = -\frac{GM}{a_H}\Bigl(1+\frac{r}{a_H}\Bigr)^{-1}.
\end{equation}
Adams et al. (2006) set $a_H=R$, with $R$ the size of the clusters obtained from Lada \& Lada (2003). The total mass can be
converted to a 'central density' through $\rho_0=M/(2\pi a_H^3)$. The total mass inside $a_H$ is $\frac{1}{4}M$ and here $\rho(r) \sim
r^{-1}$. In order to model clusters where $\rho(r) \sim r^0$ close to the centre, one could use a Dehnen profile with $\gamma=0$.
However, we decided to settle for the density profile of Plummer (1911), which is widely used in star cluster simulations because of
its simplicity (e.g. Aarseth et al. 1974; Baumgardt \& Kroupa, 2007; Kroupa et al., 2001). Its density profile is given
by

\begin{equation}
 \rho_P(r) = \frac{3M}{4\pi a_P^3}\Bigl(1+\frac{r^2}{a_P^2}\Bigr)^{-5/2},
\end{equation}
where the central density is $\rho_0 = 3M/(4\pi a_P^3)$ and the density at the centre scales as $\rho(r) \sim r^0$. Here $a_P$ is the
Plummer radius. The potential is given by

\begin{equation}
 \Phi_P(r)=-\frac{GM}{a_P}\Bigl(1+\frac{r^2}{a_P^2}\Bigr)^{-1/2}.
\end{equation}
Here we shall use both the Hernquist and Plummer distributions only for their simplicity and ability to adequately reproduce
the observed density structure in the centre of the cluster.\\

An additional quantity to address is the magnitude of the velocity dispersion within the clusters. Observations indicate that in { the
youngest} embedded clusters the { velocities of starless clumps and young stellar objects are a fraction of the virial value. Thus,} the
orbits of stars are mostly radial. Hydrodynamical simulations of cluster formation from dynamically hot gas results in { the formation of
young stellar objects that have speeds comparable to the sound speed, which are much lower than the virial value} (Bate et al., 2003). An
example of a cluster with sub-virial { speeds} is L1688, part of the $\rho$ Oph complex, where the { velocities of the stars are}
approximately 30\% of the virial value (Andr\'{e} et al., 2007). Similar results are found in $\rho$ Oph A (Di Francesco et al., 2004) at
50\% of the virial value, while NGC 2264 (Peretto et al., 2006) and NGC 1333 (Walsh et al., 2007) have even lower values. Indeed, some
{ of the pre-stellar cumps and some young stellar objects in these clusters} appear to exhibit a collapse because of the { low
speeds}. { The deviation from virial equilibrium of a cluster is quantified by the parameter $Q$, which is the ratio between the
total kinetic energy and total potential energy; virial systems have $Q=\frac{1}{2}$.} For our purpose we shall consider a velocity
dispersion with a value between 0.3 and 0.5 of the virial value { i.e. $Q \in [0.05, 0.125]$}.\\

The last issue we discuss is mass segregation. Massive stars are believed to sink to the cluster center over a relatively long
time scale, given approximately by $t_R/M_*$, where $t_R$ is the dynamical relaxation time, and $M_*$ is the mass of the star in solar
masses (Portegies Zwart, 2009). Of course, the massive stars can also be formed at the cluster centres, and some observational evidence
(Testi et al. 2000; Peretto et al., 2006) and theoretical considerations (Bonnell \& Davies 1998; McKee \& Tan, 2003) support this
point of view. However, Allison et al. (2009) report very rapid mass segregation if the initial structure of the cluster is
sufficiently fractal and the velocities are highly subvirial, typical for these young clusters. In their study of these highly fractal
clusters with subvirial velocities, mass segregation was achieved for stars heavier than 5~$M_{\odot}$ and completed in approximately
1~Myr. This rapid mass segregation appears consistent with observations of the Orion Nebula Cluster (Hillenbrand \& Hartmann, 1998;
Moeckel \& Bonnell, 2009). In this study we consider clusters which are already mass segregated for stars heavier than 5~$M_{\odot}$.
It should be noted that the rapid mass segregation coincides with a violent relaxation phase after the gravitational collapse, both of
which are a result of its subvirial velocities, as the cluster tries to reach equipartition. As it does so it shrinks in size by
approximately a factor two or more (Allison et al., 2009). In this study we do not concern ourselves with the mechanism behind mass
segregation, whether it is through dynamics or by formation, but assume it has already happened for stars heavier than
5~$M_{\odot}$. \\

Now that we have discussed most of the properties of embedded clusters based on the latest observations, and constrained some of the
key parameters that will be used, we next describe our numerical methods.

\section{Initial conditions and numerical methods}
In this section we describe the initial conditions and methods employed for our numerical simulations.

\subsection{Initial conditions and gas removal}
We generate the stars in each cluster as follows. First, the desired number of stars was chosen. The mass of each star was
then calculated randomly according to the Initial Mass Function formulation of Kroupa et al. (1993), with the functional form

\begin{equation}
 M(\xi) = 0.08 + \frac{0.19\xi^{1.55}+0.05\xi^{0.6}}{(1-\xi)^{0.58}}\, M_{\odot}.
\end{equation}
Here $\xi$ is a number chosen randomly on the interval $[0,1)$ and $M(\xi)$ is the mass of the star in solar masses. The average
stellar mass is then $\langle m_* \rangle =\int M(\xi) d\xi \approx 0.43$~$M_{\odot}$. The total mass of the stars is approximately
$M_* = \langle m_* \rangle N$, and is valid for large $N$. No primordial binaries were included. { In order to model the Solar System we
generated one star with a mass of exactly 1~$M_{\odot}$ and considered it to be the Sun.} The tidal radius of the cluster was computed as
$r_t = 1.92\,(N/100)^{1/3}$~pc, and the core radius is either $\frac{1}{6}r_t$ or $\frac{1}{3}r_t$. For the Plummer profile the Plummer
radius, $a_P$, is then computed from $r_c$ by using $\sigma_0 = \bigl(\frac{GM}{6a_P}\bigr)^{1/2}$ and solving for $a_P$, resulting in $a_P
= \sqrt{2}r_c$. For the Hernquist profile, the central velocity dispersion is 0, so we use $\rho(r_c)=\frac{1}{3}\rho_0$ and solve for $a_H$
to find $a_H \approx 1.46r_c$.\\

The stars in the cluster are subjected to three forces. The first is their mutual gravitational interaction. The second is caused by
the Galactic tide and bulge, and is modelled according to the formulation of Levison et al. (2001) but with the Oort constants set at
$A=|B|=15$~km~s$^{-1}$~kpc$^{-1}$ respectively (MacMillan \& Binney, 2010). The third force is caused by the gas that is present in the
cluster, whose density profile is also modelled either by the Hernquist or Plummer distribution, with the same values { of $a_H$
or $a_P$ used for the stars and gas}. The mass in gas is related to the total mass in stars, $M_*$ and the star formation efficiency (sfe,
$\varepsilon$) by $M_g=(\varepsilon^{-1}-1)M_*$. The sfe was set either to 0.1 or to 0.25, which mostly brackets the observed range
(Lada \& Lada, 2003). \\

The magnitude of the position and velocity vectors of the stars are generated from the isotropic energy distribution functions of either the
Hernquist or Plummer profiles using a von Neumann rejection technique (Press et al., 1992). For the Plummer model we followed Aarseth et al.
(1974). They solve the equation $M(r)=\xi M$ for $r$ for each star, with $\xi=nM_*/N$ and $n<N$. This results in successive values of $M(r)$
being evenly spaced, which appears to be fine for the heavy stars that have settled in the centre, but is artificial for the other stars.
Thus we decided to adopt this method for the heavy, segregated stars but replaced $\xi$ by a random number on the interval between
$[0,\xi_{\rm max})$ for the other stars. Here $\xi_{\rm max}$ is determined by $M(r)=M(r_t)$. For the Hernquist model we followed the same
procedure. The singularity of the distribution function of the Hernquist model at { energy} $E=-GM/a_H$ is avoided by noting that $4\pi
r^2 v^2 f(|E|) < 1.1$, where $f(E)$ is the Hernquist distribution function and $v$ is the velocity of a { star. We use the latter}
formulation with the von Neumann technique (D.~C.~Heggie, personal communication). The position and velocity vectors were calculated with
random orientation. In order to avoid the Sun leaving the cluster immediately, we 'rigged' the system by requiring that the Sun moves
inwards if it is farther then $\frac{1}{2}r_t$ from the centre when the simulation is started. This was accomplished by requiring that for
the Sun $\vec{r}\cdot \vec{v} <0$ if $r>\frac{1}{2}r_t$. \\

After generating the velocities and positions from the distribution function the kinetic energy of the stars was reduced. { A single
value for the virial parameter $Q$ for each simulation was randomly computed on the interval $Q \in [0.05, 0.125]$, and the kinetic energy
of each star was then reduced accordingly.} The maximum speed of the stars is $v=(2Q)^{1/2}v_{\rm{esc}}$ where
$v_{\rm{esc}}=|2\Phi(r)|^{1/2}$ is the local escape speed and $\Phi(r)$ is the gravitational potential of the Hernquist or Plummer sphere.
The stars' low kinetic energy implies that the stars are initially on nearly radial orbits and exhibit almost a free fall (which justifies
having the Sun move inwards at first). The resulting set of stellar positions and velocities was then used as the starting conditions in our
full simulations. \\

In the above procedure, all stars were assumed to have formed at the same time. There is an ongoing debate about whether or not
this is true, but based on observations of NGC 2264 Peretto et al. (2006) favour the turbulent formation mechanism of McKee \& Tan
(2003), in which most stars form within $\sim 10^5$~yr, with the heaviest stars in the centre before the initial collapse of the
cluster. Palla \& Stahler (1999) report a similar conclusion based on the Orion cluster while Murray (2011) states that most stars form
within the free-fall time of the cluster. These results suggest that most stars in the cluster form within a short time of each other,
justifying our procedure.\\

The next thing to model is the decay of the gas. The early work by Lada et al. (1984) removed the gas either instantaneously or on a
time scale ranging from three to four crossing times. { The crossing time is defined as the time it takes for a star to cross the whole
cluster i.e. $t_c = 2R/\sigma$, where $\sigma$ is the velocity dispersion. For our clusters $t_c \sim 4$~Myr.} Adams et al. (2006) kept the
gas density constant and then removed it instantaneously after 5~Myr, independent of the properties of the cluster. Baumgardt \& Kroupa
(2007) removed the gas on a time scale from zero to ten crossing times. Proszkow \& Adams (2009) removed the gas instantaneously at times
ranging from 1~Myr to 7~Myr, to account for the spread in observed embedded cluster lifetimes. Levison et al. (2010) kept the gas density
constant for 3~Myr and removed it exponentially with an e-folding time of 10\,000~yr. Thus, { previous studies show great variability in
their choice of the time of gas removal and its decay rate.} An initially bound cluster responds to external perturbations and changes in
the potential in approximately a crossing time and thus this time scale serves as a bench mark. Kroupa (2000) reports that typically the gas
in an embedded cluster is removed on a crossing time. However, for our purposes we do not want to end up with a bound system of which the
Sun is a potential member, and thus we would like to take the gas away quickly to ensure that the Sun escapes from the system. Even though
Proszkow \& Adams (2009) took the gas away instantaneously, they frequently found that bound systems remained. The strongest dependency
appears to be on the star formation efficiency, $\varepsilon$, but even when setting $\varepsilon=0.1$, they reported that some 15\% of
systems remained bound after the gas went away. Ultimately, the gas in the cluster is removed because stellar winds from heavy stars create
H$\alpha$ bubbles and supernovae. Stellar winds cause rapid outflow of H$\alpha$ bubbles, typically with velocities of $v_w\sim
25$~km~s$^{-1}$ (Whitmore et al., 1999). This leads to a removal time scale of $t_d = R/v_w \sim 80\,000$~yr, for clusters with $N \in [10,
1000]$. Given this rapid time scale, we have decided to proceed as follows: the gas is kept at its initial density for a time of either
2~Myr or 4~Myr. Given that the typical lifetime of an embedded cluster is some 5~Myr (Lada \& Lada, 2003) with a maximum of 10~Myr, and that
circumstellar discs have a median lifetime of 3~Myr (Currie et al., 2008), and that Jupiter and Saturn had to form in this time (Lissauer \&
Stevenson, 2007), a typical cluster lifetime after the formation of Jupiter and Saturn of 1~Myr to 5~Myr is reasonable. After the constant
density phase, the density of the gas decays exponentially with an e-folding time of $t_d=30\,000$~yr.\\

\subsection{Numerical integration method}
Star clusters are not well-suited to the symplectic integration methods typically used to follow the orbits of comets about stars
(Spurzem et al., 2009). These methods split the Hamiltonian into the sum of a part representing Keplerian motion around a fixed star
and a part representing the non-Keplerian perturbations from other stars (Wisdom \& Holman, 1991). However, we find that the orbits of
the stars and other bodies are efficiently computed using the standard ‘Leap-Frog’ integrator, which is a second-order symplectic
integrator where the Hamiltonian is split into the sum of a part representing the kinetic energy and one representing the gravitational
potential energy. As described in Duncan et al. (1998), the key idea is to incorporate a multiple time step symplectic method such that
whenever two or more stars, or a comet and star(s), or comet and planet suffer close encounters, the time step for the relevant bodies
is recursively divided to whatever level is required to resolve the encounter. In other words, the code splits the encounter into a
series of 'shells' which Duncan et al. (1998) place apart as $R_i/R_{i+1}=3^{2/3}$ and the time step is divided by 3 when crossing into
the next shell. Since these requirements are already implemented in the SyMBA integration package (Duncan et al., 1998), we modified it
to use the Leap-Frog scheme. From now on this new code will be referred to as SyMBAC. For a description of tests of the code without
planets we refer to Levison et al. (2010).\\

Including the Jovian planets { orbiting the Sun} in SyMBAC proved to be tricky. The Leap-Frog integration scheme is not very accurate for
Kepler orbits unless one takes of the order of 1\,000 steps per orbit (Kokubo et al., 1998). However, even with such a small time step it
introduces a secular drift in the argument of pericentre, $\omega$, of the planets (Kokubo et al., 1998). This drifting in $\omega$ changes
the secular properties of the planets and it is easy to grow the eccentricities of Jupiter and Saturn to crossing values. One solution to
this problem is to invent a new symplectic method to handle both Kepler orbits and stellar encounters, or to search for a set
of parameters that keep the planets stable. We chose to do the latter. It turned out that it is imperative to keep Jupiter and Saturn
on the same 'shell', and that the shells are scaled as $R_i/R_{i+1}=3$ and the time step between shells is divided by 6. The first
'shell' around the Sun is set to 2916~AU and the time step is 100~yr, and is used for all of the calculations that follow. Since the
typical spacing between the stars is much larger than 3\,000~AU, encounters between stars are relatively rare and the code does not
often need to resort to the recursive encounter routines apart from in the beginning when there are many comets close to the Sun. Hence
the simulations we performed were finished in mere hours rather than days or weeks. As in BDL6 we do not include Uranus and Neptune,
because their formation mechanism is still not well understood and the time scale of their formation is not well constrained (Goldreich
et al., 2004).\\

Comets were added { only} around the Sun as bodies with masses { thirteen} orders of magnitude smaller than the stars and planets.
The stars and planets { mutually interact}, but the comets { are only affected by the stars and the planets, and not by each other}.
The comets were started on orbits between 4.5~AU and 12~AU with eccentricities and inclinations between 0 and 0.01 (radians). A total of
4\,000 comets were used in each simulation. In order to hasten the evolution, we removed comets from the simulation when they were farther
than 1.5$r_t$ from the cluster centre and unbound. { In order to keep the calculations traceable, and because we are only interested in
the formation of the inner Oort cloud with indigenous comets, no comets were added around other stars, unlike in Levison et al. (2010).} The
simulations were stopped either after some maximum time (15~Myr) or when the Sun was farther than 1.5$r_t$ from the cluster centre after the
gas began to evaporate.\\

The last parameter to choose is the number of stars in the cluster. We have chosen seven individual values which bracket the
approximate observed range of embedded clusters: 50, 100, 250, 350, 500, 750 and 1\,000. A total of 40 simulations were performed for
each cluster, with 10 for each combination of the time the gas density remained constant, $t_g$, and the concentration $c$. Together
with the two values of the sfe and the two potential/density pairs, this results in a total of 1\,120 simulations, which were all
performed on the CRIMSON Beowulf cluster at the Observatoire de la C\^{o}te d'Azur.

\subsection{Sample cluster evolution}
\begin{figure*}
\resizebox{\hsize}{!}{\includegraphics[]{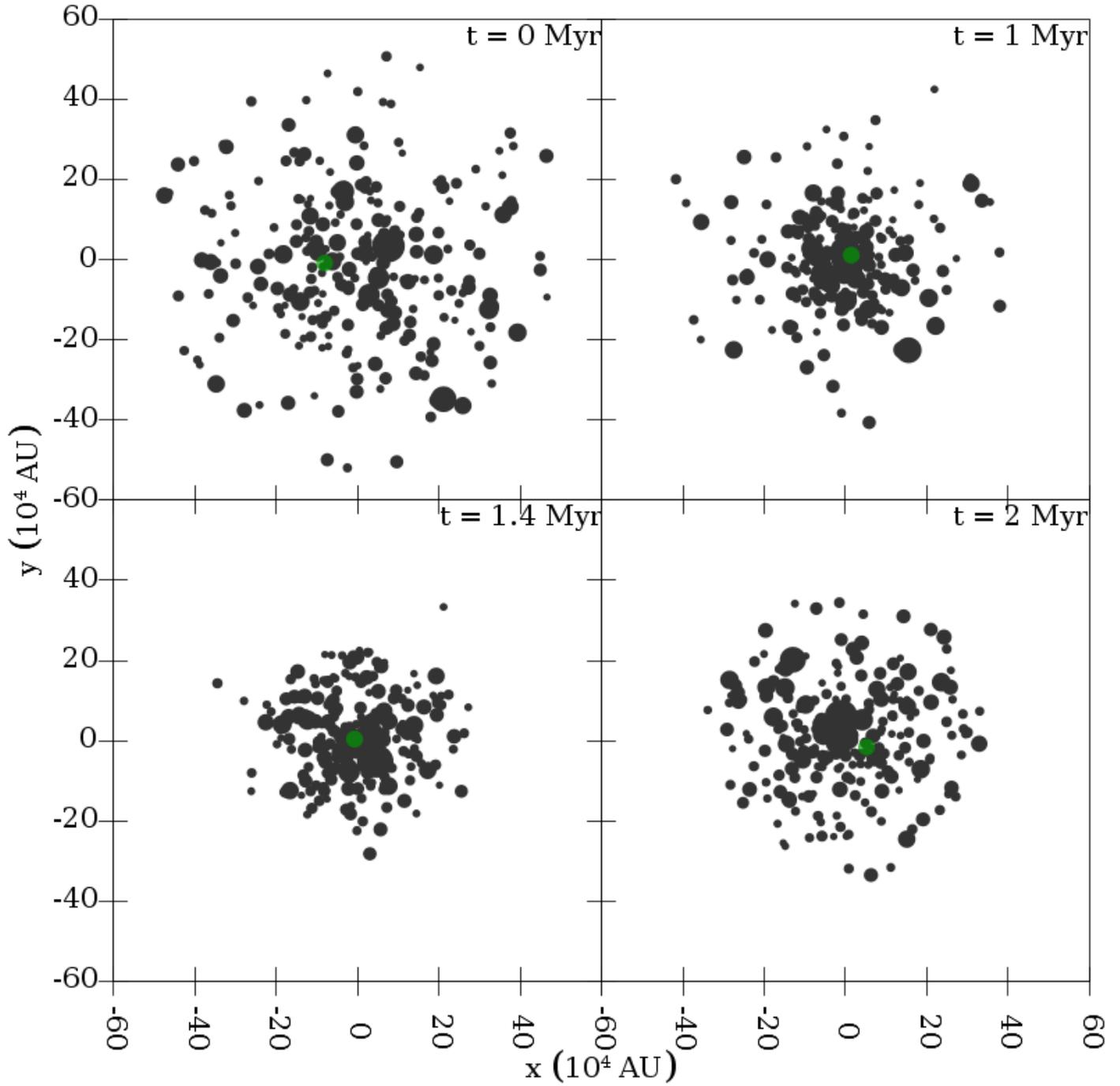}}
\caption{First 2~Myr of the dynamical evolution of a sample cluster with $N=250$ and a Hernquist distribution. The positions of the
stars are projected onto the $xy$-plane. The size of the bullets scales as the stellar mass $M_*^{1/3}$. The green bullet represents
the Sun.}
\label{cl1}
\end{figure*}

\begin{figure*}
\resizebox{\hsize}{!}{\includegraphics[]{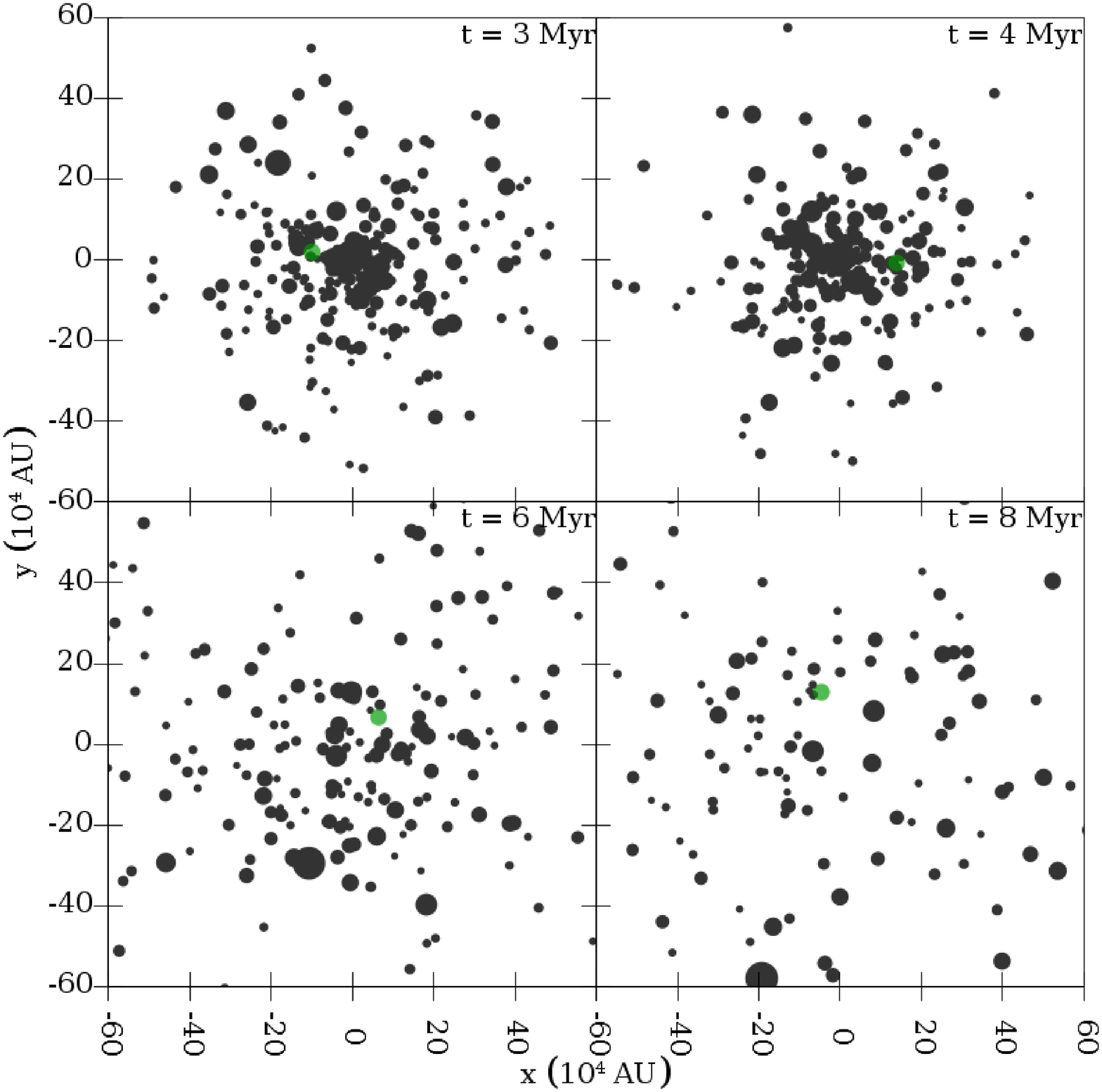}}
\caption{A continuation of Fig.~\ref{cl1}. The gas begins to decay after 4~Myr.}
\label{cl2}
\end{figure*}

In this subsection we give an example of the evolution of one of the many star clusters that we simulated. The cluster is of the
Hernquist type with $N=250$, $\varepsilon = 0.1$, $t_g=4$~Myr and the { concentration parameter} $c=6$. Figures~\ref{cl1} and~\ref{cl2}
show snapshots of the stars in the $xy$-plane at times indicated in the top-right corners of the panels. The Sun has been coloured in green
to distinguish it from the other stars, and the size of the dots correlates to the mass of the star as $M_*^{1/3}$. \\

The top-left panel of Fig.~\ref{cl1} depicts the initial positions of the stars, projected onto the $xy$-plane. These were generated
according to the methods described above. Since the stars are all sub-virial, the cluster undergoes a state of collapse, through a
process called 'violent relaxation' (Lynden-Bell, 1967). Some embedded clusters are observed to be in this state of collapse
(Peretto et al., 2006; Andr\'{e} et al., 2007). As one can see in the top-right panel of Fig.~\ref{cl1}, the cluster has
shrunk significantly and the density in the core is much higher than it was initially. After about 1.4~Myr, the cluster has reached its
smallest size and maximum core density. The cluster has a size of approximately 1~pc. After this maximum compression of the core, the
cluster expands again (bottom-right panel) and some of the stars are either unbound or on very elongated orbits. Continuing on to
Fig.~\ref{cl2} we see that the cluster continues to expand and reaches some sort of a steady-state. At 4~Myr the gas goes away and in
the bottom two panels of Fig.~\ref{cl2} we see the stars escaping from the system. This sequence of events, in which the cluster
collapses and then expands again, occurs in all our simulations.\\

Now that we have given a brief overview of the dynamical evolution of the star clusters, we turn to the results of our simulations of
stars and comets.

\section{The formation of the inner Oort cloud}
In this section we present the results of our numerical simulations. Many properties of the inner Oort cloud that are formed in this
way are similar to earlier results presented in Fern\'{a}ndez \& Brun\'{i}ni (2000), BDL6 and Kaib \& Quinn (2008), such as the the
cloud being nearly isotropic apart from the inner 10\% or so. We shall not repeat all of these here but instead only focus on the key
aspects { and how these scale with the cluster properties: the distribution in semi-major axis, inclination and perihelion, the
location of the inner edge and the formation efficiency.} During this investigation it became clear that the parameter space is very large,
so we shall try to reduce it first. When examining the output from the simulations, it turned out that the initial virial ratio, $Q$, does
not impact the results beyond the statistical noise one expects from one simulation to another. { This result is surprising, because the
value of $Q$ determines how much the cluster shrinks during the initial collapse and violent relaxation.} Thus, in what follows we consider
the results to be averaged over the value of $Q$. This leaves us with the number of stars, $N$, the star-formation efficiency,
$\varepsilon$, the time the gas was removed, $t_g$, the concentration, $c$, and the density profile (Plummer or Hernquist). For reference, a
comet is defined to be in the cloud if it satisfies both $a>50$~AU and $q > 35$~AU. \\

\subsection{The size and the concentration radius of the inner Oort cloud}
\begin{figure}
\resizebox{\hsize}{!}{\includegraphics[angle=-90]{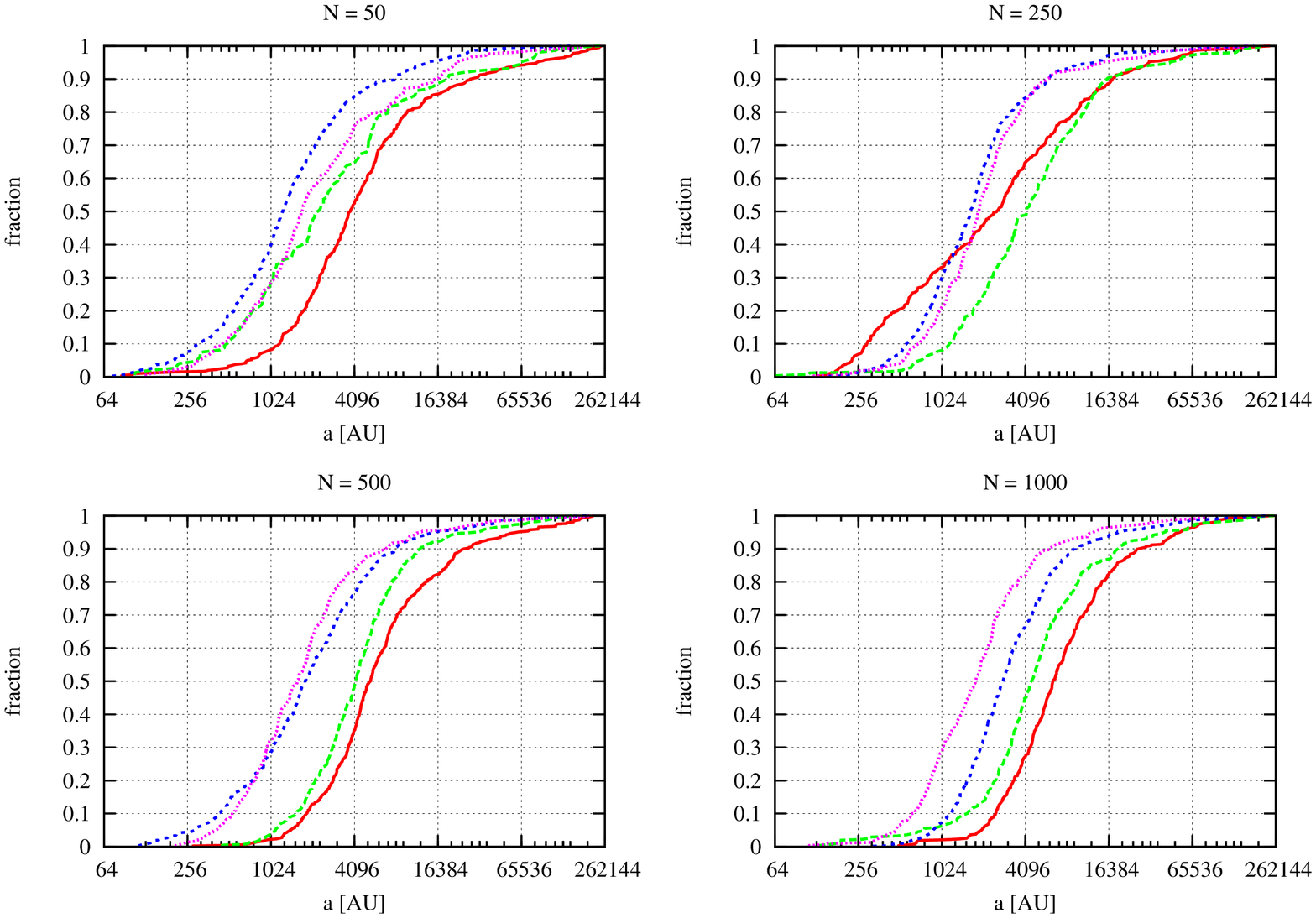}}
\caption{Cumulative semi-major axis for Oort clouds for various Hernquist clusters. Red line: $t_g=2$~Myr, $r_t=3r_c$. Green:
$t_g=4$~Myr, $r_t=3r_c$. Blue: $t_g=2$~Myr, $r_t=6r_c$. Magenta: $t_g = 4$~Myr, $r_t=6r_c$. Data is for sfe of 10\%.}
\label{cumah}
\end{figure}

\begin{figure}
\resizebox{\hsize}{!}{\includegraphics[angle=-90]{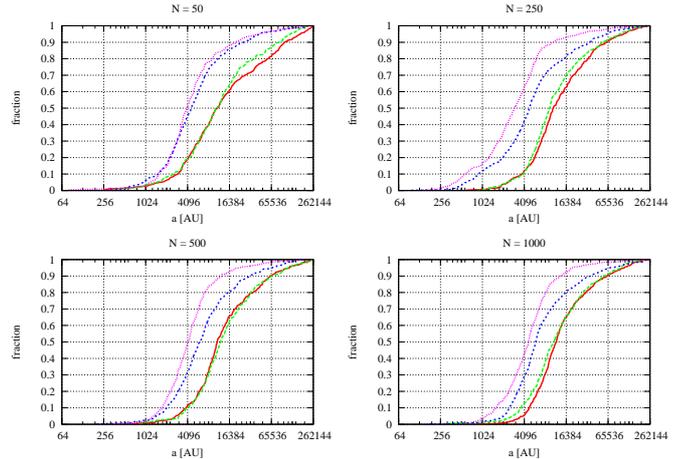}}
\caption{Same as Fig.~\ref{cumah} but now for Plummer { clusters}. Once again the sfe is 10\%.}
\label{cumap}
\end{figure}

A simple way to depict the size and distance of the inner Oort cloud to the Sun is to plot the cumulative semi-major axis distribution
of comets that are in the cloud. We have plotted this distribution for Hernquist clusters with 50, 250, 500 and 1\,000 stars
(Fig.~\ref{cumah}) and Plummer (Fig.~\ref{cumap}) distributions respectively. Both figures correspond to clusters with an sfe of 10\%.
The red lines are for clusters with $t_g=2$~Myr and $c=3$, green lines have $t_g=4$~Myr and $c=3$, the blue lines correspond to
the parameters $t_g=2$~Myr with $c=6$, while the magenta lines depict cases with $t_g=4$~Myr and $c=6$. We will use these
$t_g-c$-colour correlations throughout the paper, unless specified otherwise. The plots show a few interesting features that require
further discussion. \\

The first is that, generally but not exclusively, the magenta lines precede the blue lines, which precede the green lines, which in
turn precede the red lines. In other words, the inner Oort clouds formed above become more centrally condensed when both $t_g$ and $c$
increase. This is easy to imagine. As the Sun resides in the cluster for longer times, the outermost comets are stripped away
by encounters with the other stars. Only the tightest-bound comets survive these encounters over long times because stars need to
come ever closer to destabilise the close-in comets. In addition, as the Sun spends more time in the cluster, the tidal forces from the
gas and stars have more time to torque the comets' perihelia away from the planets. Since this time scale is $t_q \propto
\rho_g^{-1}a^{-2}$ (Duncan et al., 1987), spending more time in the cluster will torque comets at smaller semi-major axis. The same
argument holds for the torquing of the comets' perihelia by stellar encounters. Secondly, ignoring the red curve in the upper-right
panel of Fig.~\ref{cumah}, where the Sun suffered a very close encounter { with another star} early on, the range of the curves of the
same colour are similar among all the panels of each model. However, when comparing the Hernquist and Plummer clouds, for the Hernquist
model the range of the clouds is from approximately 200~AU to 200\,000~AU (where we truncated it), while for Plummer the clouds are farther
away, starting at about 800~AU. This makes sense: the central density in the Hernquist model is higher than that in the Plummer model and
thus as the Sun flies through the centre of the cluster the torquing by the gas and the encounters with the other stars in the
Hernquist model are more violent than in the Plummer model. A third interesting feature is that the cumulative distributions { have a
similar shape and appear just different in their median values. It appears as if both stochastic effects from one simulation to the next
and the number of stars in the cluster affects the final distribution. Kaib \& Quinn (2008) also reported that their results suffered from
strong stochastic effects from one simulation to the next.} We examined the cumulative semi-major axis distributions for clusters with an
sfe of 0.25 and conclude that they are very similar to the figures shown above.\\

\begin{figure}
\resizebox{\hsize}{!}{\includegraphics[angle=-90]{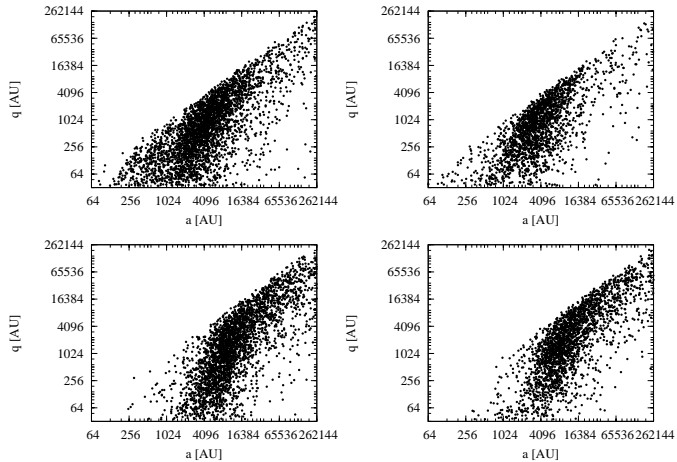}}
\caption{A few sample Oort clouds in semi-major axis-pericentre space. The top panels are for Hernquist clusters, the bottom two are for
Plummer models.}
\label{ocfig}
\end{figure}

In order to characterise the Oort clouds better, in Fig.~\ref{ocfig} we have plotted a few sample Oort clouds in semi-major axis-pericentre
space. The data are accumulated over a series of clusters with different $N$. The top two panels pertain to Hernquist models while the
bottom two panels are for Plummer clusters. Note that the Plummer clusters are less centrally concentrated than the Hernquist clusters.
Note also that unlike the traditional Oort cloud, which has an inner edge at approximately 2000~AU (e.g. Dones et al., 2004, Kaib \& Quinn,
2008) the Oort cloud formed in the cluster environment can extend all the way down to $\sim$200~AU so that the cluster environment
produces an Oort cloud that is much more centrally condensed.\\

The differences among the cumulative distributions are a measure of the concentration of the cloud. Unfortunately there is no definite
way to define this from the cumulative distributions of the semi-major axis. However, we might turn to the theory of star clusters for
guidance. In analogy with star clusters, a useful way to characterise the central concentration of the inner Oort cloud is by
considering the number density of the comets as a function of semi-major axis. This is similar to counting the stars in a cluster as a
function of their projected distance to the centre. The number density of the comets as a function of the semi-major axis, $n(a)$, can
be well approximated by a power law when far from the Sun, and the slope is usually -$\frac{7}{2}$ or $-4$ (Duncan et al., 1987; Dones
et al., 2004; BDL6; Kaib \& Quinn, 2008), while close to the Sun the density is usually flat (e.g. Brasser et al., 2010). This suggests
a best fit through the number density profile of the form $n(a)=n_0(1+a/a_0)^{-4}$, where $n_0=3N_c/(4\pi a_0^3)$ (Dehnen, 1993;
Tremaine et al., 1994) measures the central number density of the cloud, $N_c$ is the total number of comets in the cloud and $a_0$ is
a parameter that measures the central concentration. The lower its value, the more centrally condensed the cloud is, and probably the
more centrally condensed or long-lived the Sun's birth cluster was. The number of comets as a function of semi-major axis is then
$N_c(a) = N_c(\frac{a}{a+a_0})^3$, which is a reasonable approximation to the curves presented in Figs.~\ref{cumah} and~\ref{cumap}.
The 'half-mass' radius $r_{1/2} = (2^{1/3}-1)^{-1}a_0 \approx 3.85a_0$. In addition $N_c(a_0)=\frac{1}{8}N_c$, so that most of the
comets are in the 'halo' rather than in the core. \\

\begin{figure}
\resizebox{\hsize}{!}{\includegraphics[angle=-90]{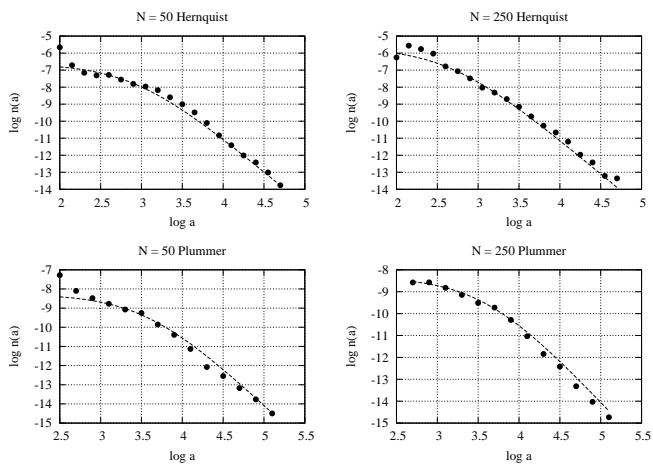}}
\caption{Density profiles corresponding to the red lines in upper panels of Figs~\ref{cumah} and~\ref{cumap}. The bullets show the
data, the lines are the best fits of the form $n(a)=n_0(1+a/a_0)^{-4}$.}
\label{den1}
\end{figure}

We have performed a series of fits to the data to characterise the range and typical value of $a_0$. While performing the fits, we
encountered cases where a better expression was $n(a)=n_0(1+a^2/a_0^2)^{-2}$, where $n_0=N_c/(\pi^2a_0^3)$ (Elson et al., 1987) and
$N_c(a)=(2N_c/\pi)\tan^{-1}(a/a_0) - (2N_c/\pi)(a/a_0)(1+a^2/a_0^2)^{-1}$. This distribution turns quicker from $a^0$ to $a^{-4}$ around
$a_0$ than the Dehnen distribution. However, it is slightly more compressed, because the half-mass radius is at $r_{1/2}\approx2.26a_0$,
though $N_c(a) = 0.182\,N_c$. As an example, for the red curve in the top-right panel in Fig.~\ref{cumah}, in which the inner Oort cloud is
rather concentrated towards the centre, we have $a_0=445$~AU, but for the red curve in the upper-left panel $a_0=825$~AU. For the same
curves using Plummer profiles, $a_0 = 3\,631$~AU and $a_0 = 3\,825$~AU respectively. The density profiles for these examples, and their best
fits, are given in Fig.~\ref{den1}. We find large variations for $a_0$ as a function of $N$ for one set of values of $t_g$ and
$c$. The value of $a_0$ depends more sensitively on $c$ than on $t_g$, and decreases as either of these increase. { We found no
correlation between the value of $a_0$ and $Q$, suggesting that the previous two parameters have a stronger effect on the final structure
of the inner Oort cloud.} For reference, for the classical Oort cloud that was formed in the current Galactic environment (Dones et al.,
2004; Kaib \& Quinn, 2008) $a_0 \sim 5000$~AU. \\

\begin{figure}
\resizebox{\hsize}{!}{\includegraphics[angle=-90]{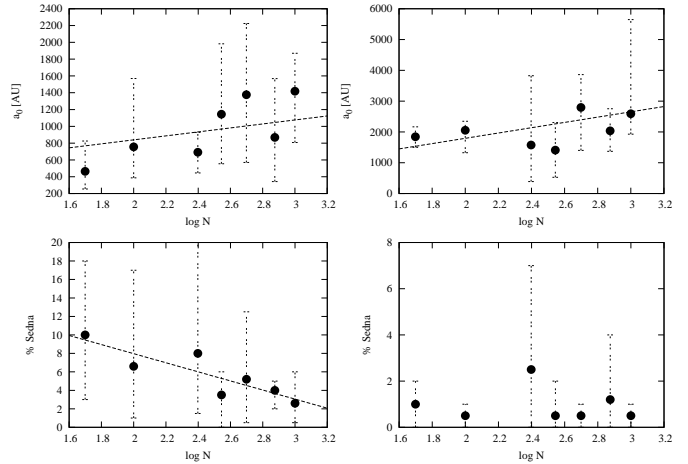}}
\caption{Top: Average value of the scaling parameter $a_0$ as a function of cluster membership $N$. Bottom: Location of Sedna in the
cumulative semi-major axis distribution. The lines show the best linear fit through the data. The sfe is 10\%. Error bars denote the
maximum and minimum values for each quantity as a function of $N$. Left column: Hernquist. Right column: Plummer.}
\label{seff}
\end{figure}

\begin{figure}
\resizebox{\hsize}{!}{\includegraphics[angle=-90]{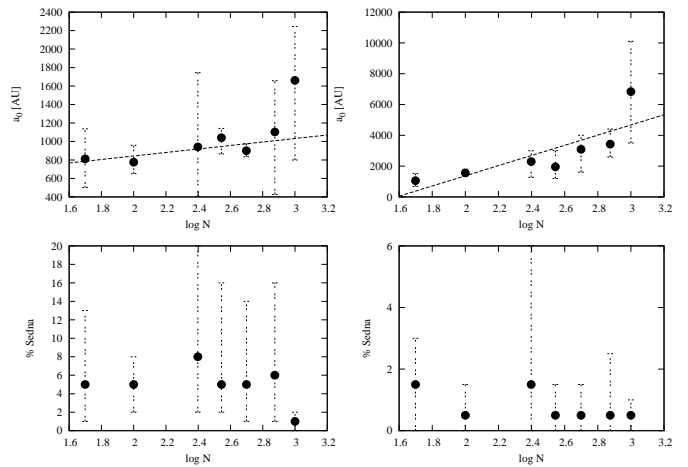}}
\caption{Same as Fig.~\ref{seff} but now the sfe is 25\%.}
\label{seff2}
\end{figure}

We have plotted the values of $a_0$, averaged over the various combinations of $t_g$ and $c$ as a function of $\log N$ in the top two
panels of Fig.~\ref{seff} for an sfe of 10\%, and in Fig.~\ref{seff2} for clusters with sfe of 25\%. Error bars denote the maximum and
minimum values that we obtained from our data for each value of $N$ and are a proxy for the values of $t_g$, $c$ { (and $Q$)} and
how stochastic { the dynamics is. The error bars indicate that our previous assessment of the variations the cumulative
semi-major axis distribution being due to stochastic effects was essentially correct.} The left panels refer to Hernquist clusters while the
right panels refer to Plummer models. As one can see, there is a trend for $a_0$ to increase with $N$. Best fits through the data yield a
steeper slope for Plummer than for Hernquist in both cases. Even though the fitting errors are large, there {\it is} a trend for $a_0$ to
increase with $N$. This trend appears to be systematic and we get back to it later. \\

\subsection{Sedna}
In BDL6 we concluded that the median distance of the inner Oort cloud to the Sun scales with the cluster density as $a_{50} \propto
\langle \rho \rangle^{-1/2}$. In BDL6 we said that this combination of $a$ and $\rho$ is found in the torquing time of the
pericentre, $t_q$, and thus if the product $a\rho^{1/2}$ is constant, then $t_q$ must be a constant. From BDL6 we find for the
inner edge $a_i = 7\,700(\rho_0/100\,M_{\odot}\,\rm{pc}^{-3})^{-1/2}$ AU. Thus, to get (90377) Sedna, an inner Oort cloud dwarf planet
with $a \sim 500$~AU (Brown et al., 2004), we need a density of $\rho \sim 20\,000$~$M_{\odot}$~pc$^{-3}$. Do we see this occur in our
clusters and does it occur for long enough to torque objects onto Sedna-like orbits? \\

In the bottom panels of Fig.~\ref{seff} we plot the value of the cumulative semi-major axis distribution at Sedna's location
($a=500$~AU), $f_s$, as a function of $\log N$. The bottom-left panel once again refers to Hernquist clusters while the bottom-right
panel are the Plummer models. The data is averaged over $t_g$, $c$ { and $Q$}, which are incorporated into the error bars. Once again
these denote the maximum and minimum values that we observed. For the Hernquist case the data can be fit as $f_s = 17.81\% -
4.92\%\log N$, with errors of 17\% and 25\% on the coefficients. For Plummer and $\varepsilon = 0.1$ a Sedna is found in only 70\% of
the cases, but then only typically at the 1\% level of the cloud, with no trend in $N$. When averaging over all $N$, the average
values of $f_s$ for Hernquist are 4.8\% for $(t_g,c)=(2,3)$, 7.7\% for (2,6), 2.7\% for (4,3) and 7.6\% for (4,6), so that $f_s$
depends more sensitively on $c$ than on $t_g$. { We report that in most cases $f_s$ increases as either $c$ or $t_g$ increases, but we
found no systematic correlation between $f_s$ and $Q$.} \\

The trend of $f_s$ decreasing and $a_0$ increasing with $N$ can be explained as follows. In order to obtain a Sedna, the Sun needs to (i)
temporarily pass through an environment with very high density, (ii) pass through that part of the cluster where the torquing on the comet
is at a maximum, or (iii) experience a close stellar passage. From BDL6 we know that in a Plummer cluster the torquing on the comet is a
maximum when $r_{\odot} \sim 0.8a_P$. Unfortunately, this does not coincide with the maximum density, and indeed the torque on a comet
vanishes at the centre of the cluster. Thus, to produce Sedna in a Plummer cluster requires a close stellar passage. In contrast, for the
Hernquist model both the density and the torque are a maximum at the centre of the cluster, so that the only requirement is that the Sun
passes close to the centre. However, the possible trajectories that the Sun can have to get close to the centre and experience the spike in
density and torque that are needed, decrease with increasing $N$ because the size of the cluster itself increases. In other words, the orbit
of the Sun needs to be more and more radial with increasing $N$ and thus the initial conditions for these orbits occupy a smaller region of
phase space for the larger clusters than for smaller ones. The figures above indicate that the location and existence of Sedna { are only
mildly dependent on the cluster parameters}.\\

\subsection{Inclination and perihelion distribution}
\begin{figure}
\resizebox{\hsize}{!}{\includegraphics[angle=-90]{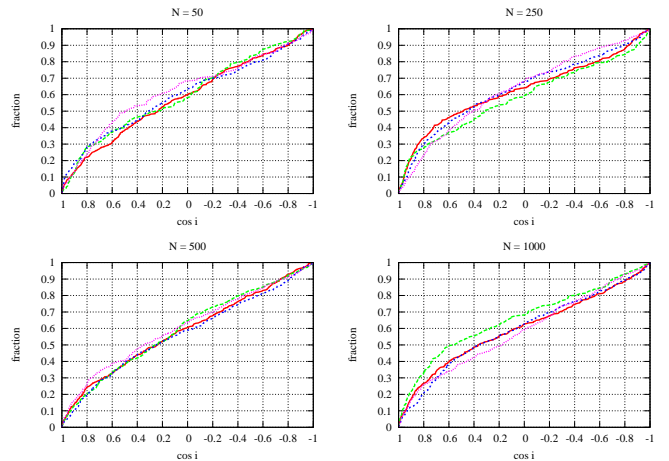}}
\caption{Cumulative inclination for Oort clouds for various Hernquist clusters. Red line: $t_g=2$~Myr, $r_t=3r_c$. Green:
$t_g=4$~Myr, $r_t=3r_c$. Blue: $t_g=2$~Myr, $r_t=6r_c$. Magenta: $t_g = 4$~Myr, $r_t=6r_c$.}
\label{cumi}
\end{figure}

\begin{figure}
\resizebox{\hsize}{!}{\includegraphics[angle=-90]{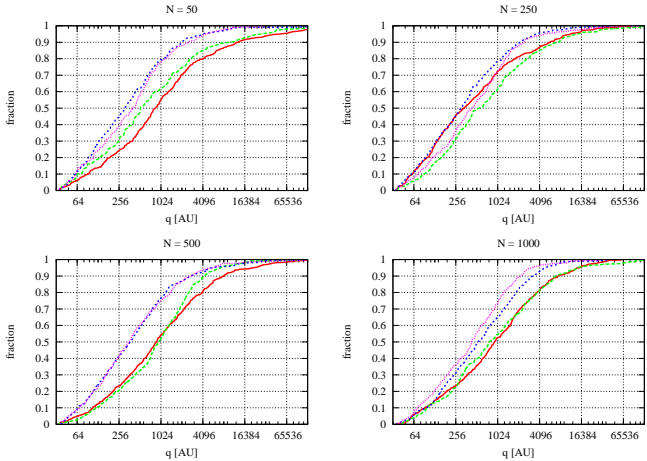}}
\caption{Same as Fig.~\ref{cumi} but now for the perihelion distance.}
\label{cumq}
\end{figure}

Figure~\ref{cumi} shows the cumulative value of the cosine of the orbital inclination { in the ecliptic plane} for various inner Oort
clouds from Hernquist clusters. As one can see, there is little variation among the distributions with either $N$, $t_g$ or $c$. All clouds
are predominantly prograde since the median value of $\cos i$ is larger than 0. If the inclination distribution were isotropic, then a
cumulative distribution of $\cos i$ would be a straight line from 0 to 1 as $\cos i$ goes from 1 to $-1$. As one can see, this is not the
case for most distributions and thus the inner Oort cloud is not (yet) isotropic. That said, the median inclination is lower in the inner
parts of the cloud and higher in the outermost parts (not shown). However, even in the innermost part (less than 10\% in cumulative
semi-major axis) the median value of the inclination is typically between 45$^\circ$ and 55$^\circ$. Sedna, with it's orbital inclination of
12$^\circ$, is mostly in the bottom 5\% of the inclination distribution. Figure~\ref{cumq} plots the cumulative perihelion distribution for
the same clusters. Once again the different curves are similar in each plot. The cumulative distribution in $q$ scales fairly { well}
with $\ln q$, so that $dN/dq \propto q^{-1}$ and the differential distribution is flat in $q^{-1}$. This implies that most comets are found
with small perihelion relative to the semi-major axis. This is no surprise for two reasons. First the Galactic tide has not had time to
randomise the $q$ distribution. Second, the innermost part of the Oort cloud i.e. comets with $a \lesssim 2\,000$~AU, are barely affected by
the Galactic tides even on Gyr time scales (Dones et al., 2004; BDL6), so we expect both the perihelion and inclination distributions to be
preserved in this region. \\

When examining the inclination and perihelion distributions for the Plummer clusters, they turn out to be very similar to those for the
Hernquist clusters. This is { to be expected} because these distributions have not been evolved to the present age { and thus
the Galactic tide has not had time to mix the inclinations and eccentricities of the comets.} Regarding the cumulative eccentricity
distribution, which is not shown, we find { the cumulative distribution scales as} $f(e) \propto e^\beta$, with $\beta \in (2,3)$. A
thermal distribution has the cumulative $f(e) = e^2$. The Spearman rank correlation coefficient between the semi-major axis and eccentricity
distributions varies from -0.07 to 0.05, suggesting no trend of the eccentricity to either increase or decrease with increasing semi-major
axis. Between the semi-major axis and inclination distributions, the Spearman rank correlation coefficient is approximately 0.25, suggesting
a weak trend for the inclination to increase with increasing semi-major axis. We verified this by noting that the inner portion of the Oort
cloud has a lower median inclination than the outer part.

\subsection{The fossilised inner Oort cloud}
In BDL6 we argued that comets with semi-major axis $a \lesssim 2\,000$~AU are barely affected by the Galactic tide because the
average rotation rate of their apses is governed by planetary perturbations rather than the Galactic tide. Therefore, it is likely
that this population has preserved its original orbital structure since its formation during the time the Sun was in a cluster. We call
this population the 'fossilised inner Oort cloud', and its most prominent member is Sedna, with 2000 CR$_{105}$ { --
$(a,q,i)=(220,44,22^\circ)$ --} and 2004 VN$_{112}$ { -- $(a,q,i)=(350,47,26^\circ)$ --} being potential other members. The
high-inclination objects (136199) Eris and 2004 XR$_{190}$ (Buffy) are excluded because their existence can be explained by other dynamical
mechanisms { (Gomes, 2011)}. Here we are interested in the orbital distribution of objects in this fossilised inner Oort cloud in order
to directly compare it with observations. However, before proceeding there are a few issues that we have to consider.\\

Recently Schwamb et al. (2010) attempted to constrain some of the properties of the Sun's birth cluster by comparing recent deep-field
observations with Palomar of trans-Neptunian objects with the simulations presented in BDL6. Schwamb et al. (2010) concluded
that the current environment of the outer Solar System is incompatible with the two densest clusters from BDL6 with a confidence of
95\% or better; they were unable to reject the next-densest clusters, with central density $\rho_0 = 10\,000$~$M_{\odot}$~pc$^{-3}$,
for which Sedna was always at the inner edge. This should help us put some crude constraints on the cluster models employed in the
current study that are compatible with the data from Schwamb et al. (2010) and which are not. Re-examining the data from BDL6 we find
that for the $10^4$~$M_{\odot}$~pc$^{-3}$ clusters Sedna is located at the 3\% level in cumulative semi-major axis, and the innermost
object has a semi-major axis of 218~AU. For the denser clusters the inner edge is closer in. However, how can we distinguish between objects
that were placed on their current orbits by a stellar encounter and by planetary actions, in particular during the late epoch of planetary
instability where Neptune most likely temporarily resided on a highly eccentric orbit (Tsiganis et al., 2005)? Gomes et al. (2005) { and
Gomes (2011)} have demonstrated that the combined effect of mean-motion resonances with Neptune and the Kozai mechanism (Kozai, 1962) can
place objects from the Scattered Disc on orbits with a high inclination (up to 50$^\circ$) and large pericentre distance (more than 40~AU),
mimicking a fossilised inner Oort cloud object that was perturbed by a star. In addition, as Neptune migrated and its eccentricity decreased
through dynamical friction, some objects detached from is resonances or sphere of influence and now permanently reside on high-perihelion,
high-inclination orbits (Gomes et al., 2005; { Gomes, 2011)}. However, the combined effects of mean-motion resonances and the Kozai
mechanism breaks down when the mean-motion resonances with Neptune begin to overlap, which occurs for semi-major axes larger than 200 to
250~AU (Gomes et al., 2005; Lykawka \& Mukai, 2007). Thus any object with a semi-major axis longer than this value, and a pericentre
distance farther than 38~AU -- beyond which chaotic diffusion stops (Gladman et al., 2002) -- could be a fossilised inner Oort cloud object.
Therefore, for our purpose, we eliminate objects with semi-major axis shorter than 250~AU and perihelia lower than 38~AU. \\

\begin{figure}
\resizebox{\hsize}{!}{\includegraphics[angle=-90]{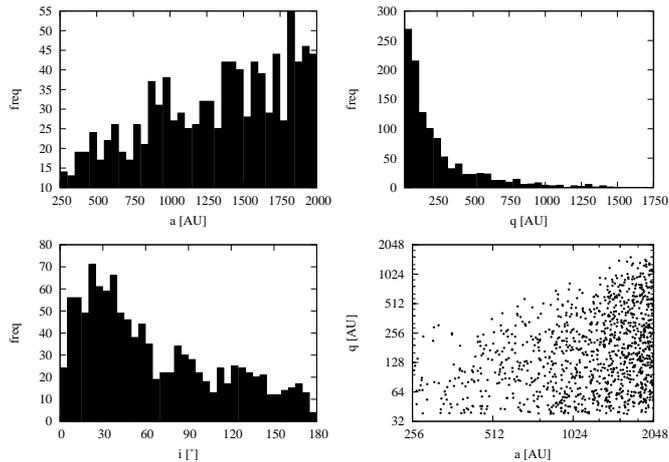}}
\caption{Several properties of a fossilised inner Oort cloud. This plot uses all the data from the Plummer model with sfe of 10\%.
The top-left panel plots { a histogram of the $a$ distribution}. The top-right panel shows a histogram of the $q$ distribution. The
bottom-left panel shows a histogram of the inclination distribution while the bottom-right panel depicts { $a$ vs $q$}. All distributions
are normalised.}
\label{fossilp}
\end{figure}

The cumulative distributions of the inclination and pericentre distance for the fossilised inner Oort cloud are very similar to those
depicted { in Figs.~\ref{cumi} and~\ref{cumq}}above, apart that the distribution in $q$ rolls over earlier than above. We decided instead
to combine all the data from the Plummer simulations with sfe of 10\% into one plot. In Fig.~\ref{fossilp} we plot, in the top-left panel, a
histogram of the semi-major axis distribution for all the objects in the fossilised inner Oort cloud. The stochastic behaviour of the
semi-major axis distribution is caused by the chaotic nature of the simulations and the peaks and troughs should not be considered as
definitive, although the increasing trend is a systematic effect. The top-right panel depicts a histogram of the $q$ distribution, and
matches the $q^{-1}$ profile discussed earlier i.e. most objects are trapped with a low value of $q$, with the median value being $\sim
150$~AU, but $\sim 200$~AU for the Hernquist clusters. The bottom-left panel displays the inclination distribution { clearly indicating
the prograde bias}, while the bottom-right panel plots the semi-major axis vs pericentre distance.

\subsection{Efficiency}
\begin{figure}
\resizebox{\hsize}{!}{\includegraphics[angle=-90]{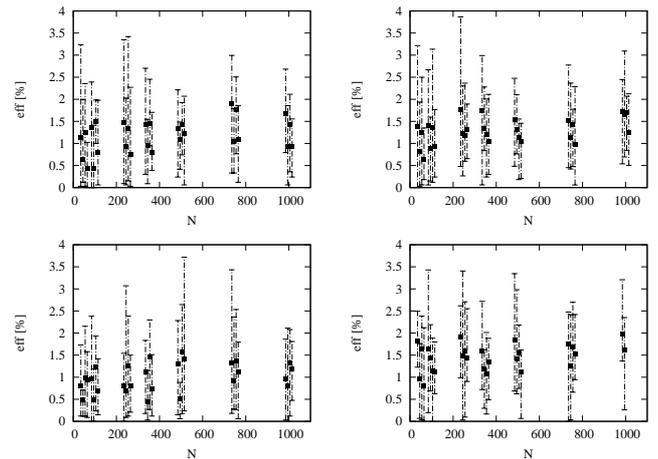}}
\caption{Formation efficiency of the Oort cloud as a function of $N$. Each filled square lists the average efficiency and the error
bars denote the maximum and minimum. Each filled square is for a different combination of $t_g$ and $c$, which are, from left to right,
(2,3), (2,6), (4,3) and (4,6). The left panels are Hernquist clusters with sfe 10\% (top) and 25\% (bottom). The right panels are for
Plummer clusters with the top having an sfe of 10\% and the bottom 25\%.}
\label{feff}
\end{figure}

Here we report the formation efficiency of the inner Oort cloud from our numerical simulations. We have plotted the formation
efficiency as a function of the number of stars in the cluster in Fig.~\ref{feff}. Each filled square denotes the average efficiency
for a certain combination of $t_g$ and $c$ and the error bars denote the maximum and minimum. The combinations of $t_g$ and $c$ are,
from left to right, (2,3), (2,6), (4,3) and (4,6). The left panels are for Hernquist clusters with an sfe of 10\% (top) and 25\%
(bottom). The right panels are for Plummer clusters with the top having an sfe of 10\% and the bottom 25\%. As is clear, the typical
efficiency does not depend strongly on the properties of the cluster but remains at approximately 1.5\%. BDL6 reported typical
efficiencies of 10\% although Kaib \& Quinn (2008) obtained efficiencies of typically 3\% during their open cluster phase, which are in
better agreement with our current value. In addition, the results of Kaib \& Quinn (2008) do not strongly depend on their cluster
properties either. We cannot pinpoint the exact reason for the significant difference in the trapping efficiency between our current
simulations and those presented in BDL6 although we can present a plausible argument. \\

Kaib \& Quinn (2008) argued that their resulting inner Oort clouds and efficiencies were very strongly dependent on the closest stellar
passage through their Oort clouds. If this passage happened late then there would be little material left in the Oort cloud or
the Scattered Disc to refill the cloud (Levison et al., 2004). In addition, in BDL6 the orbit of the Sun remained fixed in the Plummer
potential and were obtained from a cluster in virial equilibrium, so that most orbits had a small radial excursion. Since the Sun
stayed near the Plummer radius the fluctuations in density and flux of stellar encounters that the Sun experienced were limited. Here
this is not the case: the Sun's orbit is nearly radial so that it passes close to the cluster centre before receding back into the halo
after the initial phase of violent relaxation. Thus the Sun does not stay close to the Plummer or Hernquist radius and, especially in
the latter clusters, experiences a change in background density of at least an order of magnitude. The passages through the centre of
cluster occur after 1 to 3 Myr, by which time the number of comets that are being scattered by Jupiter and Saturn has decreased
by as much as 80\%. The close encounters with massive stars near the centre of the cluster and the decrease in tidal radius around the
Sun strip many of its comets when it passes through the centre of the cluster. By then there are not many comets left to resupply the
Oort cloud and thus the corresponding efficiency is low. This efficiency could possibly be increased by adding Uranus and Neptune and
having them scatter the comets in their vicinity. However their formation mechanism and time scale is not well understood (Goldreich et
al., 2004). Thus we have reason to believe that the efficiencies that we obtained here are in agreement with the expected
dynamics. \\

Is the existence of Sedna in agreement with a typical trapping efficiency of 1.5\%? Levison et al. (2008) and Morbidelli et al. (2009)
estimate that there were approximately 1\,000 Pluto-sized objects in the trans-Neptunian disc. Since the mass in this disc is probably
comparable to that in the Jupiter-Saturn region, we also estimate that there were some 1\,000 Pluto-sized objects in this region.
Taking the size distribution for large Kuiper belt objects from Bernstein et al. (2004), with a cumulative slope of 3.5, we estimate
there were approximately 4\,000 Sedna-sized objects shortly after the formation of Jupiter and Saturn. If only 1.5\% of these ended up
in the Oort cloud, this implies there are some 60 Sedna-sized objects in the Oort cloud, and thus we can expect to see one of these in
the innermost 2\% of the cloud. Thus, our low formation efficiency is compatible with the detection of one Sedna-like object. However,
this estimate is uncertain by factors of a few because of the lack of knowledge of the primordial number of Sedna-like objects in the
Jupiter-Saturn region.\\

In the next section we { better compare} the compatibility of our cluster models with a single detection of Sedna.

\section{Observational Constraints on the Sedna population}
In this section we compare the orbital distribution of { Sedna-like} bodies produced in the above cluster environments to the
observational constraints the wide-field survey of Schwamb et al. (2010) (hereafter S10 ) place on the Sedna population. S10 searched
11\,786 deg$^2$ down to a mean limiting $R$ magnitude of $\sim$21.3, within $\pm$30$^\circ$ of the ecliptic. With the exception of Sedna, no
new Sedna-like bodies with perihelia beyond 45~AU were found despite a sensitivity out to distances of $\sim$1000 AU. For further details
about the survey, we refer the reader to S10. Here we apply the same technique and survey simulator, developed in S10, to test whether the
new cluster orbital distributions can serve as the source of the Sedna population and are consistent with the single re-detection of Sedna.
For each cluster environment, we compare the orbital distributions of single detections produced by the survey simulator to Sedna.
We employ a modified 3-dimensional Kolmogorov-Smirnov (3-D KS) test, detailed in S10, which simultaneously compares the semi-major axis,
inclination, and eccentricity ($a$, $e$, $i$) distributions of single detections produced by each cluster environment to Sedna's orbital
parameters ($a$=519 AU, $e$=0.853, $i$=11.9$^\circ$). \\

From each cluster-created orbital distribution, we randomly generate 3 million orbits from the final inner Oort clouds from our simulations,
obtaining the semi-major axis, inclination, and eccentricity for each inner Oort cloud candidate object. The S10 observations probe the
present-day inner Oort cloud population, in particular the Sedna region, but the orbital distributions presented here are valid only shortly
after the cluster dissipates and the Sun exits the cluster. Although Sedna's orbit and other objects in the fossilised inner Oort cloud are
dynamically protected from the effects of passing stars and galactic tides in the current solar environment, we must account for the 4.6~Gyr
of subsequent evolution the Solar System has undergone and account for those effects that may have sculpted other objects in the Sedna
region since emplacement. To avoid confusion with Scattered Disc and detached Kuiper belt objects, we conservatively exclude objects with
perihelia smaller than 50~AU and semi-major axes shorter than 250~AU because these could have dynamically interacted with Neptune. We also
exclude orbits with semi-major axes greater than 3\,000~AU because these objects are capable of becoming long period comets (Kaib \& Quinn
2009). BDL06 obtain a value of $\sim$ 2~Gyr for the precession frequency of Sedna; therefore we assume that the orbits of other
Sedna-like objects have been randomised due to planetary effects, and thus randomly choose all other orbital angles. \\

To create our sample of single detections, we randomly assign absolute magnitudes to each of our simulated inner Oort cloud populations. Due
to the large uncertainties in the albedos of such a distant population, we choose to assign absolute magnitudes rather than sizes. We
assume a single power-law brightness distribution where the number of objects brighter than or equal to a given absolute magnitude,
$H_{\rm{max}}$, is described by:
\begin{equation}
 {\rm N} (H \le H_{\rm{max}})=N_{H\leq1.6}10^{ \alpha (H_{\rm{max}}-1.6)}
 \end{equation}
The brightness distribution is scaled to $N_{H\leq1.6}$, the number of bodies with an absolute magnitude brighter than or equal to Sedna
($H=1.6$). \\

For both the Hernquist and Plummer cluster models we performed the 3-D KS test for a possible range of values for $\alpha$ (0.2-0.82), 
including $\alpha=0.35$ and $\alpha=0.82$, the best-fit values for the hot ($i>5^\circ$) and cold ($i<5^\circ$) KBOs respectively
(Fraser et al., 2010). Each single instance of the brightness distribution can be thought of as a separate survey, and we continued
sampling the brightness distribution until we created 10\,000 synthetic single detections of inner Oort cloud objects for every cluster
environment and each value of $\alpha$. We found that for any value of $\alpha$, the 3-D K-S probability of rejection is a relatively
flat value as $N_{H\leq1.6}$ increases. Therefore we restrict our analysis and discussion to $N_{H\leq1.6}$ =1. \\

Table \ref{tab:cluster} lists the results from the K-S tests for each cluster environment. We have sampled the clusters according to
sfe and number of stars, thereby crudely averaging over the concentration and $t_g$. For the first column the first four letters refer to
the potential that was used, either Hern or Plum, e01 or e25 refers to the sfe being either 10\% of 25\%, and n50 to n1000 refers to the
number of stars in the cluster. Except for the Plume01n100 model, which marginally produced Sedna, and the Plume25n100 model, {\it all} the
orbital distributions are consistent with the S10 observations of a single detection of a Sedna-like body, and cannot be rejected at greater
than the 90$\%$ confidence level for the entire range of $\alpha$ that was tested. The majority of the simulated cluster environments
reproduced Sedna's orbit at the inner edge of the distribution, emplacing the bulk of the population onto orbits with semi-major axes and
perihelia greater than Sedna's. { Unfortunately} we find no direct correlation between cluster size and the model fit, and the majority
of the Plummer and Hernquist potential clusters are consistent with the S10 observations. Though the synthetic Hernquist potential
populations provide a slightly better fit to the data, we are unable to reject the majority of the Plummer potential clusters. The Hernquist
distribution has a wider range of eccentricities at lower semi-major axes for the single produced single detections than for the Plummer
model. \\

{ Given that all models appear almost equally viable in the production of Sedna we must conclude} that the formation of a Sedna
population is a natural outcome of the cluster birth scenario independent of the initial conditions and properties of the cluster, { and
that the small variations are just the result of stochastic effects in the cluster}. { Until this work} the 10$^4$~$M_{\odot}$~pc$^{-3}$
cluster of BDL06 { was} the only orbital distribution consistent with the detection of a single Sedna-like body in the S10 survey. For
$\alpha=0.35$ and $\alpha=0.82$ respectively, S10 find the best-fit values for the number of objects brighter than or equal to Sedna are
595$^{+ 1949}_{-400}$ and 112$^{+ 423}_{-71}$ for the 10$^4$ $M_{\odot}$~pc$^{-3}$ cluster of BDL06. The Plummer and Hernquist distributions
in this study resemble the 10$^4$ $M_{\odot}$~pc$^{-3}$ cluster orbital distribution, with Sedna located at the inner edge of the orbital
distribution with many more objects having semi-major axes and eccentricities larger than Sedna. We expect these clusters would produce
similar population estimates to the 10$^4$ $M_{\odot}$~pc$^{-3}$ clusters of BDL06. To estimate the size of the Sedna population, we
selected a few of the orbital distributions from each potential to examine. For each given value of $\alpha$, we randomly assigned an
absolute magnitude 50\,000 times to each synthetic object our survey simulator created for every value of $N_{H\leq1.6}$. For each
$N_{H\leq1.6}$ tested, accounting for survey efficiency, the number of simulations in which, like the real survey, one object on a
Sedna-like orbit is detected, are tallied. Using the same slopes for the brightness distribution ($\alpha=0.35$ and $\alpha=0.82$), we find
similar results than S10. In other words, a cluster environment could emplace on the order of a hundred to a thousand planetoids brighter
than or equal to Sedna beyond the Kuiper belt, roughly consistent with the earlier estimate of approximately 100 objects based on the
efficiency of inner Oort cloud formation and a crude estimate as to the number of available objects. An order of magnitude or two more mass
may reside in the Sedna region than exists in the present Kuiper belt.

\begin{table}
\centering
\begin{tabular}{ l c c c c c c  }
\hline
\hline
Cluster model & \multicolumn{5}{c}{$\alpha$} \\
\cline{2-6}
 & $0.2$ &  $0.35$ &  $0.4$ & $0.6$  & 0.82\\
 \hline
 \hline
Herne01n50	&	45.85	&	47.67	&	47.66	&	53.45	&	44.91	\\
Herne01n100	&	58.28	&	68.13	&	66.29	&	70.58	&	75.1	\\
Herne01n250	&	71.99	&	72.21	&	69.33	&	63.66	&	57.61	\\
Herne01n350	&	67.54	&	71.88	&	69.73	&	72.21	&	64.63	\\
Herne01n500	&	36.02	&	41.74	&	45.7	&	57.65	&	65.44	\\
Herne01n750	&	49.45	&	48.44	&	47.99	&	43.71	&	42.34	\\
Herne01n1000 	&       64.05 	&    	62.0 	& 	63.9 	& 	66.9 	& 	67.3 	\\
Herne25n50	&	42.94	&	48.61	&	51.57	&	53.62	&	52.6	\\
Herne25n100	&	31.35	&	21.59	&	17.53	&	18.21	&	17.69	\\
Herne25n250	&	85.91	&	86.07	&	82.95	&	79.91	&	82.53	\\
Herne25n350	&	55.73	&	53.67	&	58.71	&	59.94	&	75.42	\\
Herne25n500	&	67.4	&	65.2	&	64.24	&	61.72	&	55.79	\\
Herne25n750	&	69.63	&	71.63	&	68.53	&	67.72	&	67.73	\\
Herne25n1000	&	77.57	&	69.16	&	64.24	&	55.21	&	38.65	\\
\hline
\hline

Plume01n50	&	66.77	&	63.63	&	69.4	&	69.79	&	76.42	\\
Plume01n100	&	99.26	&	99.75	&	99.88	&	99.93	&	99.99	\\
Plume01n250	&	76.44	&	72.62	&	71.5	&	64.69	&	60.46	\\
Plume01n350	&	82.09	&	90.54	&	90.09	&	89.91	&	91.85	\\
Plume01n500	&	84.76	&	88.47	&	86.15	&	82.37	&	82.38	\\
Plume01n750	&	23.69	&	41.26	&	41.72	&	50.57	&	55.14	\\
Plume01n1000	&	82.23	&	77.06	&	73.69	&	72.73	&	61.82	\\
Plume25n50	&	55.35	&	53.9	&	58.14	&	58.26	&	60.33	\\
Plume25n100	&	93.49	&	96.09	&	97.25	&	97.87	&	90.93	\\
Plume25n250	&	63.9	&	62.3	&	60.99	&	53.14	&	48.11	\\
Plume25n350	&	89.61	&	82.36	&	82.03	&	77.89	&	77.64	\\
Plume25n500	&	23.45	&	22.8	&	20.12	&	19.8	&	20.2	\\
Plume25n750	&	60.89	&	47.61	&	49.53	&	47.61	&	46.23	\\
Plume25n1000 	&	81.49	&	79.0	&	77.61	&	67.55	&	57.85	\\
 \hline
\hline
\end{tabular}
\caption{3D KS test results for the cluster that produced single detections compared to Sedna's orbit. We report the confidence level we can
reject the two distributions as drawn from the same parent population.}
\label{tab:cluster}
\end{table}

\section{Discussion}
The fact that many clusters seem to fit the currently-known structure of the outer Solar System requires further elaboration. From the
above simulations that we performed and the present location of Sedna, we cannot constrain the number of stars in the Sun's birth cluster.
The only conclusion we can draw is that { almost all the clusters are consistent with the detection of one Sedna-like object at the inner
edge of the Oort cloud, and thus the production of Sedna at its current location is a generic outcome. The observational data marginally
favours the Hernquist model over the Plummer clusters, even though the former produces an inner Oort cloud that is more centrally condensed
than the latter. In order to determine which model best corresponds with the reality deeper surveys, such as LSST (Large Synoptic Survey
Telescope), which can probe the Sedna region directly, are needed}. \\

Unfortunately, our simulations remain inconclusive as to what value of $N$ best fits the current observational data. Several { previous}
studies favour a large value of $N$, typically 1\,000 to 10\,000, for reasons such as supernova enrichment, gas disc truncation through
photo-evaporation, gas disc truncation through close stellar passages and the survival of the planets in a stellar environment (Adams,
2010). On the other hand, Gounelle \& Meibom (2008) argue that it is unlikely that the Solar System formed in an environment similar
to that of the Orion Nebula Cluster (ONC), and argued instead that the Sun is a second-generation star that formed after some of the stars
in the first generation had already gone supernova. After a few Myr of evolution, star formation in ONC-like settings occurs mainly in
photo-dissociation regions where the HII region created by the central star and the surrounding molecular gas meet, which is at a few
parsecs from the massive star. The fate of the ONC is well illustrated by the 2-3 Myr old cluster NGC 2244, whose most massive star has
the same spectral type as $\theta^1$ C Ori. In NGC 2244, star formation is occurring in the outskirts of the cluster, at distances
5-10~pc from the central O6 star. Gounelle \& Meibom (2008) argue that the Sun formed in such a secondary environment and that some of
the short-lived radio nuclei that are found in meteorites were inherited from the interstellar medium rather than supernova
enrichment, or by irradiation. The first generation cluster had to be large in order for it to have undergone several supernovae, but
constraints on the second-generation are weaker. Unfortunately, at this stage we are unable to model the second-generation cluster
from which the Sun could have formed, and instead rely on the first generation, from which we cannot make a definitive conclusion
about the size and membership of the Sun's cluster. We cannot rule out that the Sun formed in a cluster with $N \gtrsim 1\,000$ stars as
advocated by some studies, but at the same time we cannot rule out clusters with $N \sim 100$ either. Most stars appear to form in
clusters with $N \in (100, 1\,000)$ stars (Lada \& Lada, 2003; Adams et al., 2006), all of which are dynamically consistent with the
observed orbital distribution of the outer Solar System.\\

Given the results above the natural question to ask is whether or not we could have done things differently. We performed a
comprehensive search of the literature to determine what are the best initial conditions and properties for the embedded clusters and
the gas expulsion. The only thing that we can think of is that our models do not update the mass distribution of the gas as the
cluster evolves, and neither does it account for a central cavity in the gas distribution around the heaviest star(s). Both of these
could affect the dynamics of the cluster. To our knowledge no N-body simulations of star clusters containing gas take into account the
motion of and the related changes in the gas. Doing so would most likely require a hybrid study of N-body for the stars, planets and
comets, and hydrodynamics for the gas. This would be a completely new study in star cluster dynamics far beyond the scope of
this paper. \\

A second issue concerns the initial conditions of the cluster. For this study we chose to artificially have the stars reach their
final masses at the same time, while in reality this occurs over a certain time span. The sequential star formation could prevent the
early violent relaxation phase of the cluster and thus the dynamics would be milder. Adams et al. (2006) incorporated the sequential
star formation by adding the stars one by one to his simulations over a time span of 1~Myr. Levison et al. (2010) randomised the phases
of the stars by integrating the whole cluster in the Hernquist potential for 1~Myr. But, as we argued earlier, there is no definitive
piece of evidence that supports either sequential star formation or having most of the stars in the cluster form at more or less the
same time. Further study is needed to support either mechanism and how it would impact the outer parts of the solar system.\\

One additional issue that we need to address is the problem of when is $t=0$? In our simulations, $t=0$ corresponds to when Jupiter
and Saturn have fully formed and begin to scatter the comets in their vicinity. Here we assume that their time scale of formation is
short compared to the free fall time of the Sun in the cluster. The free fall time is given by $t_{\rm{ff}} = (G\rho)^{-1/2} =
1.5\,(100\,M_{\odot}\,{\rm{pc}}^{-3}/\rho)^{1/2}$~Myr. A typical number density in the centre of the clusters is $n_* = 65$~pc$^{-3}$
(Carpenter, 2000), which implies $\rho \sim 28$~$M_{\odot}$~pc$^{-3}$. Taking the sfe to be 0.1 we have $t_{\rm{ff}} \sim 1$~Myr, which is
more or less the time when the cluster reaches it maximum density in Fig.~\ref{cl1}. The formation of Jupiter and Saturn is thought to
have taken between 1~Myr and 3~Myr (Lissauer \& Stevenson, 2007) and observations indicate that circumstellar gas discs have a typical
lifetime of 3~Myr (Currie et al., 2008). Thus, it is possible that the Sun had already passed through the centre of the cluster at
least once before the formation of Jupiter and Saturn had been completed. Would including Jupiter, Saturn and the comets after this
initial passage dramatically change the outcome of our simulations? We do not think so because the Sun will still have a radial orbit
and continue to experience a few passages through the cluster centre before the gas is blown away by stellar winds. Given that the
first passage deposits most of the comets in the cloud and is also the most damaging, we believe that the outcome will be similar to
what we have presented here. \\

The last issue we would like to address is whether or not our model predicts more Sednas. As we discussed in Section~5, the existence of
Sedna cannot rule out the current cluster model. To our knowledge, the LSST will be able to place constraints on the model. By extrapolating
expected detections from our models to a survey such as LSST with a limiting magnitude of 24, and covering $\pm 20^\circ$ from the ecliptic
allows us to constrain the slope of the size distribution of objects in the Sedna region. If there are more objects in this region brighter
than or equal to Sedna, then we expect there to be of the order of several to a few hundred more bodies of similar size, similar to our
predictions from Section 5. However, the final outcome depends on the slope of the size distribution. If the slope is steep then we expect
there to be many more similar objects in this region, but if the size distribution is rather shallow then maybe only a single or handful of
objects detectable. The next-generation surveys such as LSST could hopefully shed some light on this issue.

\section{Summary and conclusions}
We have performed a series of numerical simulations of the formation of the inner Oort cloud in an embedded cluster environment. The
initial conditions and other relevant parameters of the clusters are chosen according to the most recent observations. Specifically, we
use two potential/density pairs for the cluster, those of Plummer (1911) and Hernquist (1990), two values for the central concentration
of the cluster, two values for the star formation efficiency and the time the gas is removed, and a range of values of the initial
subvirial kinetic energy of the stars. The number of stars in the clusters are 50, 100, 250, 350, 500, 750 and 1\,000. The formation of the
inner Oort cloud is modelled by having Jupiter and Saturn scatter comets in their nearby vicinity. Uranus and Neptune were not included at
this stage. We conclude the following: 

\begin{itemize}
 \item The inner edge of the Oort cloud ranges from approximately 100~AU to a couple hundred AU, while the outer edge is
beyond 10\,000~AU. Both are virtually independent on the number of stars in the cluster.
\item The central concentration of the Oort cloud is measured by the concentration radius, $a_0$. Half of the mass resides within
$\sim 3a_0 $. For Hernquist clusters $a_0$ ranges from approximately 600~AU to 1500~AU, while for Plummer clusters it ranges from
1500~AU to 4000~AU. This difference is no surprise because the Hernquist clusters are more centrally condensed. For the current Oort
cloud which formed in the current Galactic environment the concentration radius is approximately 5000~AU.
\item The concentration radius increases with increasing number of stars in the cluster, so that more populated clusters form less
compact inner Oort clouds. The reason for this is the smaller volume of phase space that the Sun can occupy to pass close enough to the
centre to experience the necessary high density and torquing to deposit the comets in the cloud.
\item The location of the dwarf planet Sedna at 500~AU is usually within the inner 5\% of the cloud for Hernquist clusters, and within
the innermost 2\% of the cloud for Plummer clusters, in agreement with the idea that it is at the inner edge of the inner Oort cloud.
\item Almost all the orbital distributions of the clusters are consistent with a single detection of a Sedna-like body, and cannot be
rejected at greater than the 90$\%$ confidence level for the entire range of the size distribution slopes that were tested. In other words,
from the position of Sedna and a comparison with the current structure of the outer Solar System we cannot constrain the number
of stars in the cluster.
\item The inner Oort clouds formed in these clusters are not isotropic, but show a slight prograde preference with a median inclination
of approximately 50$^\circ$. The inner part is significantly more prograde while the outer part is slightly retrograde.
\item The perihelion distribution is flat in $q^{-1}$, so that most inner Oort cloud objects have small perihelia compared to their
semi-major axis.
\item We examine the `fossilised inner Oort cloud', which is the region inside approximately 2\,000~AU where the Galactic tide has
barely altered the orbital elements of the comets. We find that the distributions in inclination, perihelion and semi-major axis are
similar than for the whole cloud. { In} this region we expect these distributions to be preserved { because the Galactic tide does
not randomise the inclinations and eccentricities}. The median perihelion distance is $\sim 150$~AU for Plummer and $\sim 200$~AU for
Hernquist. 
\item The typical formation efficiency of the Oort cloud is 1.5\%, lower than previous estimates, but consistent with a single detection of
Sedna.
\end{itemize}

{\footnotesize RB gratefully thanks Germany's Helmholtz Alliance and their 'Planetary Evolution and Life' programme for financial
support. MJD acknowledges funding from Canada's NSERC. HFL gratefully acknowledges funding from NASA through their Origin of Solar
Systems programme. MES and MEB acknowledge receiving funding from NASA Origins of Solar Systems Program grant NNG05GI02G. MES is supported
by a NASA Earth and Space Science Fellowship and is supported by an NSF Astronomy and Astrophysics Postdoctoral Fellowship under award
AST-1003258.}

\section{Bibliography}
{\footnotesize
Aarseth, S.~J., Henon, M., Wielen, R.\ 1974.\ A comparison of numerical methods for the study of star cluster dynamics.\ Astronomy and
Astrophysics 37, 183-187. \\
Adams, F.~C., Proszkow, E.~M., Fatuzzo, M., Myers, P.C., 2006. Early evolution of stellar groups and clusters: Environmental effects on
forming planetary systems. Astrophys. J. 641, 504-525.\\
Adams, F.~C.\ 2010.\ The Birth Environment of the Solar System.\ Annual Review of Astronomy and Astrophysics 48, 47-85.\\
Allen, L., Megeath, S.~T., Gutermuth, R., Myers, P.~C., Wolk, S., Adams, F.~C., Muzerolle, J., Young, E., Pipher, J.~L.\ 2007.\ The
Structure and Evolution of Young Stellar Clusters.\ Protostars and Planets V 361-376. \\
Allison, R.~J., Goodwin, S.~P., Parker, R.~J., de Grijs, R., Portegies Zwart, S.~F., Kouwenhoven, M.~B.~N.\ 2009.\ Dynamical Mass
Segregation on a Very Short Timescale.\ The Astrophysical Journal 700, L99-L103. \\
Andr{\'e}, P., Belloche, A., Motte, F., Peretto, N.\ 2007.\ The initial conditions of star formation in the Ophiuchus main cloud:
Kinematics of the protocluster condensations.\ Astronomy and Astrophysics 472, 519-535. \\
Bastian, N., Ercolano, B., Gieles, M., Rosolowsky, E., Scheepmaker, R.~A., Gutermuth, R., Efremov, Y.\ 2007.\ Hierarchical star
formation in M33: fundamental properties of the star-forming regions.\ Monthly Notices of the Royal Astronomical Society 379,
1302-1312. \\
Bastian, N., Gieles, M., Ercolano, B., Gutermuth, R.\ 2009.\ The spatial evolution of stellar structures in the Large Magellanic
Cloud.\ Monthly Notices of the Royal Astronomical Society 392, 868-878. \\
Bate, M.~R., Bonnell, I.~A., Bromm, V.\ 2003.\ The formation of a star cluster: predicting the properties of stars and brown dwarfs.\
Monthly Notices of the Royal Astronomical Society 339, 577-599. \\
Baumgardt, H., Kroupa, P.\ 2007.\ A comprehensive set of simulations studying the influence of gas expulsion on star cluster
evolution.\ Monthly Notices of the Royal Astronomical Society 380, 1589-1598. \\
Bernstein, G.~M., Trilling, D.~E., Allen, R.~L., Brown, M.~E., Holman, M., Malhotra, R.\ 2004.\ The Size Distribution of
Trans-Neptunian Bodies.\ The Astronomical Journal 128, 1364-1390. \\
Binney, J., Tremaine, S., 1987. Galactic Dynamics. Princeton Univ. Press, Princeton, NJ, USA.\\
Blitz, L., Fukui, Y., Kawamura, A., Leroy, A., Mizuno, N., Rosolowsky, E.\ 2007.\ Giant Molecular Clouds in Local Group Galaxies.\
Protostars and Planets V 81-96. \\
Bonatto, C., Santos, J.~F.~C., Jr., Bica, E.\ 2006.\ Mass functions and structure of the young open cluster NGC 6611.\ Astronomy and
Astrophysics 445, 567-577.\\
Bonnell, I.~A., Davies, M.~B.\ 1998.\ Mass segregation in young stellar clusters.\ Monthly Notices of the Royal Astronomical Society
295, 691-698. \\
Brasser, R., Duncan, M.~J., Levison, H.~F.\ 2006.\ Embedded star clusters and the formation of the Oort Cloud.\ Icarus 184, 59-82. \\
Brasser, R., Higuchi, A., Kaib, N.\ 2010.\ Oort cloud formation at various Galactic distances.\ Astronomy and Astrophysics 516,
A72-A84.\\
Brown, M.E., Trujillo, C., Rabinowitz, D., 2004. Discovery of a candidate inner Oort Cloud planetoid. Astrophys. J. 617, 645-649.\\
Carpenter, J.~M.\ 2000.\ 2MASS Observations of the Perseus, Orion A, Orion B, and Monoceros R2 Molecular Clouds.\ The Astronomical
Journal 120, 3139-3161. \\
Cartwright, A., Whitworth, A.~P.\ 2004.\ The statistical analysis of star clusters.\ Monthly Notices of the Royal Astronomical Society
348, 589-598. \\
Casertano, S., Hut, P.\ 1985.\ Core radius and density measurements in N-body experiments Connections with theoretical and
observational definitions.\ The Astrophysical Journal 298, 80-94. \\
Currie, T., Plavchan, P., Kenyon, S.~J.\ 2008.\ A Spitzer Study of Debris Disks in the Young Nearby Cluster NGC 2232: Icy Planets Are
Common around 1.5-3 $M_{\odot}$ Stars.\ The Astrophysical Journal 688, 597-615. \\
Dehnen, W.\ 1993.\ A Family of Potential-Density Pairs for Spherical Galaxies and Bulges.\ Monthly Notices 
of the Royal Astronomical Society 265, 250-256. \\
Dones, L., Weissman, P.~R., Levison, H.~F., Duncan, M.~J., 2004. Oort Cloud formation and dynamics. In: Johnstone, D., Adams, F.C.,
Lin, D.N.C., Neufeld, D.A., Ostriker, E.C. (Eds.), ASPC, vol. 324. ASP, San Francisco, USA.\\
Duncan, M., Quinn, T., Tremaine, S., 1987. The formation and extent of the Solar System comet cloud. Astron. J. 94, 1330-1338.\\
Duncan, M.~J., Levison, H.~F., Lee, M.~H.\ 1998.\ A Multiple Time Step Symplectic Algorithm for Integrating Close Encounters.\ The
Astronomical Journal 116, 2067-2077. \\
Eggers, S., 1999. Cometary Dynamics During the Formation of the Solar System, Ph.D. thesis, Mathematisch-Naturwissenschaftlichen
Fakult\"{a}ten der Georg-August-Universit\"{a}t, G\"{o}ttingen, Germany. \\
Elson, R.~A.~W., Fall, S.~M., Freeman, K.~C.\ 1987.\ The structure of young star clusters in the Large Magellanic Cloud.\ The
Astrophysical Journal 323, 54-78. \\
Fern\'{a}ndez, J.~A., Brun\'{i}ni, A., 2000. The buildup of a tightly bound comet cloud around an early Sun immersed in a dense
galactic environment: Numerical experiments. Icarus 145, 580-590.\\
Di Francesco, J., Andr{\'e}, P., Myers, P.~C.\ 2004.\ Quiescent Dense Gas in Protostellar Clusters: The Ophiuchus A Core.\ The
Astrophysical Journal 617, 425-438. \\
Fraser, W.~C., Brown, M.~E., Schwamb, M.~E.\ 2010.\ The luminosity function of the hot and cold Kuiper belt populations.\ Icarus 210,
944-955. \\
Gladman, B., Holman, M., Grav, T., Kavelaars, J., Nicholson, P., Aksnes, K., Petit, J.-M., 2002. Evidence of an extended scattered
disk. Icarus 157, 269–279.\\
Goldreich, P., Lithwick, Y., Sari, R.\ 2004.\ Planet Formation by Coagulation: A Focus on Uranus and Neptune.\ Annual Review of
Astronomy and Astrophysics 42, 549-601. \\
Gomes, R.~S., Gallardo, T., Fern\'{a}ndez, J.~A., Brun\'{i}ni, A., 2005. On the origin of the high-perihelion scattered disk: The role
of the Kozai mechanism and mean-motion resonances. Celest. Mech. Dynam. Astron. 91, 109-129.\\
Gounelle, M., Meibom, A.\ 2008.\ The Origin of Short-lived Radionuclides and the Astrophysical Environment of Solar System Formation.\
The Astrophysical Journal 680, 781-792. \\
Graham, R.~L. and Hell, P., 1985. On the history of the minimum spanning tree problem. Ann. Hist. Comput. 7, 43-57.\\
Guthermuth, R.~A., Megeath, S.~T., Pipher, J.~L., Williams, J.~P., Allen, L.~E., Myers, P.~C., Raines, S.~N., 2005. The initial
conﬁguration of young stellar clusters: A K-band number counts analysis of the surface density of stars. Astrophys. J. 632, 397-420.\\
Gutermuth, R.~A., Megeath, S.~T., Myers, P.~C., Allen, L.~E., Pipher, J.~L., Fazio, G.~G.\ 2009.\ A Spitzer Survey of Young Stellar
Clusters Within One Kiloparsec of the Sun: Cluster Core Extraction and Basic Structural Analysis.\ The Astrophysical Journal Supplement
Series 184, 18-83. \\
Hernquist, L.\ 1990.\ An analytical model for spherical galaxies and bulges.\ The Astrophysical Journal 356, 359-364. \\
Hillenbrand, L.~A., Hartmann, L.~W.\ 1998.\ A Preliminary Study of the Orion Nebula Cluster Structure and Dynamics.\ The Astrophysical
Journal 492, 540-553.\\ 
Innanen, K.~A., Harris, W.~E., Webbink, R.~F.\ 1983.\ Globular cluster orbits and the galactic mass distribution.\ The Astronomical
Journal 88, 338-360. \\
Jaffe, W.\ 1983.\ A simple model for the distribution of light in spherical galaxies.\ Monthly Notices of 
the Royal Astronomical Society 202, 995-999. \\
Kaib, N.~A., Quinn, T.\ 2008.\ The formation of the Oort cloud in open cluster environments.\ Icarus 197, 221-238. \\
Kenyon, S.~J., Bromley, B.~C.\ 2004.\ Stellar encounters as the origin of distant Solar System objects in highly eccentric orbits.\
Nature 432, 598-602. \\
King, I.\ 1962.\ The structure of star clusters. I. an empirical density law.\ The Astronomical Journal 67, 471-485. \\
King, I.~R.\ 1966.\ The structure of star clusters. III. Some simple dynamical models.\ The Astronomical Journal 71, 64-75. \\
Kokubo, E., Yoshinaga, K., Makino, J.\ 1998.\ On a time-symmetric Hermite integrator for planetary N-body simulation.\ Monthly Notices
of the Royal Astronomical Society 297, 1067-1072. \\
Kozai, Y., 1962. Secular perturbations of asteroids with high inclination and eccentricity. Astron. J. 67, 591–598.\\
Kroupa, P., 2000. Constraints on stellar-dynamical models of the Orion nebula cluster. New Astron. 4, 615-624. \\
Kroupa, P., Tout, C.~A., Gilmore, G., 1993. The distribution of low-mass stars in the galactic disk. Mon. Not. R. Astron. Soc. 262,
545-587.\\
Kroupa, P., Aarseth, S., Hurley, J., 2001. The formation of a bound star clus- ter: From the Orion nebula cluster to the Pleiades. Mon.
Not. R. Astron. Soc. 321, 699-712.\\
Lada, C.~J., Lada, E.~A., 2003. Embedded clusters in molecular clouds. Annu. Rev. Astron. Astrophys. 41, 57-115.\\
Lada, C.~J., Margulis, M., Dearborn, D., 1984. The formation and early dynamical evolution of bound stellar systems. Astrophys. J.
285, 141-152.\\
Larson, R.~B.\ 1985.\ Cloud fragmentation and stellar masses.\ Monthly Notices of the Royal Astronomical Society 214, 379-398. \\
Levison, H.~F., Duncan, M.~J., 1994. The long-term dynamical behavior of shortperiod comets. Icarus 108, 18-36.\\
Levison, H.~F., Dones, L., Duncan, M.~J.\ 2001.\ The Origin of Halley-Type Comets: Probing the Inner Oort Cloud.\ The Astronomical
Journal 121, 2253-2267. \\
Levison, H.~F., Morbidelli, A., Dones, L., 2004. Sculpting the Kuiper Belt by a stellar encounter: Constraints from the Oort Cloud and
scattered disk. Astron. J. 128, 2553-2563.\\
Levison, H.~F., Morbidelli, A., Vokrouhlick{\'y}, D., Bottke, W.~F.\ 2008.\ On a Scattered-Disk Origin for the 2003 EL$_{61}$
Collisional Family-An Example of the Importance of Collisions on the Dynamics of Small Bodies.\ The Astronomical Journal 136,
1079-1088. \\
Levison, H.~F., Duncan, M.~J., Brasser, R., Kaufmann, D.~E.\ 2010b.\ Capture of the Sun's Oort Cloud from Stars in Its Birth Cluster.\
Science 329, 187-190. \\
Lissauer, J.~J., Stevenson, D.~J.\ 2007.\ Formation of Giant Planets.\ Protostars and Planets V 591-606. \\
Lykawka, P.~S., Mukai, T.\ 2007.\ Resonance sticking in the scattered disk.\ Icarus 192, 238-247. \\
Lynden-Bell, D.\ 1967.\ Statistical mechanics of violent relaxation in stellar systems.\ Monthly Notices of the Royal Astronomical Society
136, 101. \\
McKee, C.~F., Tan, J.~C.\ 2003.\ The Formation of Massive Stars from Turbulent Cores.\ The Astrophysical Journal 585, 850-871. \\
McMillan, P.~J., Binney, J.~J.\ 2010.\ The uncertainty in Galactic parameters.\ Monthly Notices of the Royal Astronomical Society 402,
934-940. \\
Moeckel, N., Bonnell, I.~A.\ 2009.\ Limits on initial mass segregation in young clusters.\ Monthly Notices of the Royal Astronomical
Society 396, 1864-1874. \\
Morbidelli, A., Levison, H.~F., 2004. Scenarios for the origin of the orbits of the trans-neptunian objects
2000 CR105 and 2003 VB12 (Sedna). Astron. J. 128, 2564-2576.\\
Morbidelli, A., Levison, H.~F., Bottke, W.~F., Dones, L., Nesvorn{\'y}, D.\ 2009.\ Considerations on the magnitude distributions of the
Kuiper belt and of the Jupiter Trojans.\ Icarus 202, 310-315. \\
Murray, N.\ 2011.\ Star Formation Efficiencies and Lifetimes of Giant Molecular Clouds in the Milky Way.\ The Astrophysical Journal
729, 133-147. \\
Oort, J.H., 1950. The structure of the cloud of comets surrounding the Solar System and a hypothesis concerning its origin. Bull.
Astron. Inst. Neth. 11, 91-110.\\
Palla, F., Stahler, S.~W.\ 1999.\ Star Formation in the Orion Nebula Cluster.\ The Astrophysical Journal 525, 772-783. \\
Peretto, N., Andr{\'e}, P., Belloche, A.\ 2006.\ Probing the formation of intermediate- to high-mass stars in protoclusters. A detailed
millimeter study of the NGC 2264 clumps.\ Astronomy and Astrophysics 445, 979-998. \\
Piskunov, A.~E., Schilbach, E., Kharchenko, N.~V., R{\"o}ser, S., Scholz, R.-D.\ 2008.\ Tidal radii and masses of open clusters.\
Astronomy and Astrophysics 477, 165-172.\\
Plummer, H.~C.\ 1911.\ On the problem of distribution in globular star clusters.\ Monthly Notices of the Royal Astronomical Society 71,
460-470. \\
Portegies Zwart, S.~F.\ 2009.\ The Lost Siblings of the Sun.\ The Astrophysical Journal 696, L13-L16. \\
Portegies Zwart, S.~F., McMillan, S.~L.~W., Gieles, M.\ 2010.\ Young Massive Star Clusters.\ Annual Review of Astronomy and
Astrophysics 48, 431-493. \\
Press, W.H., Teukolsky, S.A., Vetterling, W.T., Flannery, B.P., 1992. Numerical Recipes in FORTRAN: The Art of Scientiﬁc Computing,
second ed. Cambridge Univ. Press, Cambridge, UK.\\
Proszkow, E.-M., Adams, F.~C.\ 2009.\ Dynamical Evolution of Young Embedded Clusters: A Parameter Space Survey.\ The Astrophysical
Journal Supplement Series 185, 486-510. \\
Schmeja, S., Klessen, R.~S.\ 2006.\ Evolving structures of star-forming clusters.\ Astronomy and Astrophysics 449, 151-159. \\
Schmeja, S., Kumar, M.~S.~N., Ferreira, B.\ 2008.\ The structures of embedded clusters in the Perseus, Serpens and Ophiuchus molecular
clouds.\ Monthly Notices of the Royal Astronomical Society 389, 1209-1217. \\
Schmeja, S.\ 2011.\ Identifying star clusters in a field: A comparison of different algorithms.\ Astronomische Nachrichten 332,
172-184. \\
Schwamb, M.~E., Brown, M.~E., Rabinowitz, D.~L., Ragozzine, D.\ 2010.\ Properties of the Distant Kuiper Belt: Results from the Palomar
Distant Solar System Survey.\ The Astrophysical Journal 720, 1691-1707. \\
Spurzem, R., Giersz, M., Heggie, D.~C., Lin, D.~N.~C.\ 2009.\ Dynamics of Planetary Systems in Star Clusters.\ The Astrophysical
Journal 697, 458-482. \\
Testi, L., Sargent, A.~I., Olmi, L., Onello, J.~S.\ 2000.\ Star Formation in Clusters: Early Subclustering in the Serpens Core.\ The
Astrophysical Journal 540, L53-L56. \\
Tremaine, S., Richstone, D.~O., Byun, Y.-I., Dressler, A., Faber, S.~M., Grillmair, C., Kormendy, J., Lauer, T.~R.\ 1994.\ A family of
models for spherical stellar systems.\ The Astronomical Journal 107, 634-644. \\
Tsiganis, K., Gomes, R., Morbidelli, A., Levison, H.~F.\ 2005.\ Origin of the orbital architecture of the giant planets of the Solar
System.\ Nature 435, 459-461. \\
Walsh, A.~J., Myers, P.~C., Di Francesco, J., Mohanty, S., Bourke, T.~L., Gutermuth, R., Wilner, D.\ 2007.\ A Large-Scale Survey of NGC
1333.\ The Astrophysical Journal 655, 958-972. \\
Whitmore, B.~C., Zhang, Q., Leitherer, C., Fall, S.~M., Schweizer, F., Miller, B.~W.\ 1999.\ The Luminosity Function of Young Star
Clusters in ``the Antennae'' Galaxies (NGC 4038-4039).\ The Astronomical Journal 118, 1551-1576. \\
Wisdom, J., Holman, M., 1991. Symplectic maps for the $n$-body problem. Astron. J. 102, 1528-1538.\\}
\end{document}